\documentclass[12pt]{JHEP}
\usepackage{latexsym}
\advance\hoffset by -.35in
\makeatletter
\@addtoreset{equation}{section}
\makeatother

 \newcommand{\bra}[1]{\langle{#1}|}
 \newcommand{\ket}[1]{|{#1}\rangle}
  \def\one{{\hbox{1\kern-.8mm l}}}
  \def\be{\begin{equation}}
  \def\ee{\end{equation}}
  \def\ba{\begin{array}}
  \def\ea{\end{array}}
  \def\bea{\begin{eqnarray}}
  \def\eea{\end{eqnarray}}
  
  \def\orbifold{$(-1)^{F_L} \cdot {\cal I}_4~$}
\def\orbifoldd{${\cal I}_4~$} 
 
\def\O{{\Omega}}
  \def\Z {{\sf Z\!\!Z}}

\preprint{DAMTP-2000-93\\
  MRI-P-000906}

\title{\LARGE Non-BPS Branes in a Type I Orbifold}
  
\author{Eduardo Eyras$^{~a}$ and Sudhakar Panda$^{~b}$\\

  \vskip 24pt

  ${}^a$
  {\em Department of Applied Mathematics and
  Theoretical Physics}\\
  {\em University of Cambridge}\\
  {\em Wilberforce Road, CB3 0WA Cambridge, U.K.}\\
  \email{E.Eyras@damtp.cam.ac.uk}

  \vskip 24pt

  ${}^b$
  {\em Harish Chandra Research Institute\footnote{Formerly known as 
Mehta Research Institute of Mathematics and Mathematical Physics}}\\
  {\em Chhatnag Road, Jhoosi}\\
  {\em Allahabad 211019, India}\\
  \email{Panda@mri.ernet.in}}

\abstract{
We analyse the spectrum of non-BPS branes in the type I theory on the 
orbifold $T^4/{\cal I}_4$.
We present a detailed analysis of the action of the worldsheet
parity $\Omega$ on the different D-brane boundary state sectors
of type IIB on $T^4/{\cal I}_4$, using the covariant formulation.
Using these results we derive the spectrum of branes in the type I
orbifold. We find $\Z_2$- and $\Z$- charged non-BPS branes. 
A study of the stability of these branes in the type I orbifold
is also presented. We find that the type I non-BPS D-particle 
and D-instanton remain stable in the orbifold. The D-particle carries
no charge whereas the non-BPS D-instanton can carry twisted R-R charge.}


\begin{document}
\section{Introduction}

In recent years, significant progress has been made towards
understanding the various string duality symmetries.
The validity of such dualities has often lead to new findings in
string theory. For example, the strong-weak
coupling duality between $SO(32)$ heterotic string and the type I
theory has compelled to look for the states in type I theory which
are the dual partners of stable but non-BPS states in the heterotic
string theory. A description of such states in type I theory
is found to be in terms of a tachyon kink solution on a
D-brane anti-D-brane pair \cite{Sen-0,Sen-1,Sen-2}.
Subsequently, this phenomena and other duality relations 
have given rise to a tremendous activity in
constructing and understanding the dynamics of non-BPS branes in
string theories \footnote{For reviews on non-BPS branes
see \cite{reviews}}.

The present understanding is that type IIA (IIB) theory
admits D$p$-branes which are BPS for $p$ even (odd) and non-BPS
for $p$ odd (even) for all values of $p$ ranging from zero to nine.
Furthermore, in both theories a descent relation has
been established \cite{Sen-5} involving BPS and non-BPS branes, i.e.
a D$p$-brane is seen as a descendant of a D$(p+1)$-brane.
Moreover, one can obtain a D$p$-brane of IIA (IIB)
from a D$p$-brane of IIB (IIA) by modding out with the discrete
symmetry $(-1)^{F_L}$.
BPS branes preserve half of the spacetime supersymmetries
and are stable, whereas non-BPS
branes break all the spacetime supersymmetries and are unstable.
This instability can be explained by the presence of a
tachyon in the open string spectrum of the 
brane. The effective 
action for these non-BPS branes, including the tachyon field,
has been studied in \cite{eff-action}.
Recently, a background independent effective action has been
constructed for the case of non-commutative branes 
\cite{Mukhi-CS}.

The unstable non-BPS
branes of type II theories can decay to the stable BPS branes via 
the condensation of a solitonic configuration of the tachyon field.
This tachyon condensation
has been extensively studied using string field theory
\cite{tachyon-SFT} and p-adic string theory \cite{p-adic}.
In the framework of non-commutative field theory
\cite{Gopakumar-etal} the above scenario is also understood via non-commutative
solitons \cite{noncomm}.

Although much of the above is understood in type II theories,
less is known for non-BPS branes in type I theory,
specially in orbifold backgrounds. 
The most interesting feature of the non-BPS branes in orbifold 
and orientifold theories is that they may be stable
\cite{Sen-0,Sen-1,Sen-2,Bergman-Gaberdiel-1,Witten-Ktheory,Frau-nonbps}. 
This stability appears because of the fact that the tachyon state is projected
out from the open string 
spectrum on the brane by the orbifold/orientifold symmetry.

In this paper, we study the spectrum of fractional 
and non-BPS branes in type I theory on a $T^4/{\cal I}_4$ orbifold. 
This orbifold is equivalent to 
the orientifold by the worldsheet parity-reversal $\Omega$ 
of type IIB on $T^4/{\cal I}_4$.
Consistency of this type I orbifold
implies that the theory must contain D5 and D9 branes,
with unitary and symplectic subgroups of
$U(16)\times U(16)$ \cite{I-orbifold,Ishibashi,Bianchi-1,Bianchi-2,Gimon}.

We follow the notation of \cite{Gaberdiel-Stefanski} for D-branes
where a D$p$-brane in the orbifold $T^4/{\cal I}_4$ of a type II theory
is denoted by $\ket{D(r,s)}$, $r+s=p$, where $r$ denotes the number of 
spatial Neumann directions along the fixed plane and
$s$ denotes those along the orbifolded directions\footnote{For simplicity,
for a $\ket{D(r,s)}$ and, unless otherwise stated, 
we will always choose the $s$ Neumann directions starting from 
$x^6$; so that, for instance, a $D(1,2)$
is wrapped around the directions $x^6$ and $x^7$ of the $T^4$.}.

We describe D-branes using the boundary state formalism,
where D-branes are described by physical closed string states
of the bosonic spectrum. In orbifold theories there will be 
twisted sectors as well and the D-branes may also contain boundary states
in these sectors.
Following \cite{Gaberdiel-Stefanski}, in type II theories orbifolded
by \orbifoldd (or \orbifold) one finds BPS branes that can
be either fractional or bulk branes.
Fractional branes in type IIB on $T^4/{\cal I}_4$ can exist 
for $r$ odd and $s$ even, 
and have boundary states in all sectors
of the theory\footnote{For short we use NS and R instead of
NS-NS and R-R, respectively, as subindices in the boundary states.}:
\bea
\ket{D(r,s)}_f &=& \ket{D(r,s)}_{\rm NS,U} 
\, + \, \epsilon_1 \, \ket{D(r,s)}_{\rm R,U} \\
&& + \, \epsilon_2 \, 
\sum\limits_{a=1}^{2^s} \,  {\rm e}^{i\theta_a} 
\left( \ket{D(r,s)}_{\rm NS,T_a} 
\, + \, \epsilon_1 \ket{D(r,s)}_{\rm R,T_a} \right)
\, .\nonumber
\eea
Each twisted sector is located in one of the 16 orbifold fixed planes.
Furthermore, each boundary state sector is given 
by a GSO-invariant combination of boundary states.

Bulk branes can be described as branes with only untwisted sectors:
\be
\ket{D(r,s)}_{b} =
\ket{D(r,s)}_{\rm NS,U} + \ket{D(r,s)}_{\rm R,U} \, . 
\ee
In general for type II orbifold they exist for the
same value as the fractional branes do, since two fractional branes with
opposite twisted charges can join and give rise to a bulk brane. This brane
can then move off the fixed planes.

Non-BPS branes in type II $\Z_2$ orbifolds are referred to as
truncated branes and are represented as follows:
\be
\ket{D(r,s)}_t = \ket{D(r,s)}_{\rm NS,U} 
+ \epsilon \,
\sum\limits_{a=1}^{2^s} \,  {\rm e}^{i\theta_a} \, 
 \ket{D(r,s)}_{\rm R,T} \, .
\ee
In type IIB on $T^4/\Z_2$ they exist for $r $ and $s$ odd
\cite{Gaberdiel-Stefanski}.
They carry twisted R-R charge only, so 
they are referred to as $\Z$-charge non-BPS branes. 
In the conventions of this paper, these branes have a tension given by:
\be
T_{(r,s)} = (\alpha^\prime)^{3 - p \over 2} (2\pi)^{3-p} \sqrt{\pi}
\, ,\qquad r+s=p \, ,
\label{tension-truncated}
\ee
which coincides with the tension of a BPS D$p$-brane in type IIA;
and a (twisted R-R) charge
\be
{\tilde Q}_{(r,s)} = (2\pi \sqrt{\alpha^\prime})^{3-r} \pi^{- {3 \over 2}}
(\alpha^\prime)^{-1} \, ,
\ee
which does not depend on the number of Neumann directions along the
orbifolded 4-torus.

On the other hand, BPS branes in type I theory
are given in terms of boundary states as in type II with the
addition of a 9-crosscap\footnote{Here, and in subsequent formulae,
the background D$9$ (and eventually the D$5$) 
branes are not explicitly displayed, but are understood.}:
\be
\ket{Dp} = {1 \over \sqrt{2}}
\left( \ket{Dp}_{\rm NS} + \ket{Dp}_{\rm R} + \ket{C9} \right)
\, .
\ee
The 9-crosscap is the closed string description of the
orientifold 9-plane. BPS branes exist for $p=1,5,9$, which
are the values of $p$ for which the corresponding R-R potential
survives the $\O$-projection.

Non-BPS branes in type I are given by just an NS-NS boundary state
plus the crosscap contribution:
\be
\ket{Dp} = {1 \over \sqrt{2}}
\left( \ket{Dp}_{\rm NS} + \ket{C9} \right)
\, .
\ee
For $p=-1,0,7$ and $8$ the contribution from the crosscap
is such that the tachyons in the open strings ending on these non-BPS
branes are 
projected out and thus giving the possibility of having
stable non-BPS D-branes in type I \cite{Sen-2,Witten-Ktheory,Frau-nonbps}.

The theory we consider in this article is type I on $T^4/{\cal I}_4$,
which can be seen either as an orbifold of type I or
as an orientifold of the type IIB orbifold. Therefore, the spectrum of
D-branes of the type I orbifold can be deduced by either applying
the orbifold projection on the type I D-branes or applying the 
$\Omega$ projection to the D-branes of the type IIB orbifold. For the
first case we must take into account as well the twisted sectors.
Our approach consists in first studying the $\Omega$ projection on the
boundary states of the type IIB orbifold, for which we use the covariant
formulation of the boundary 
states \cite{covariant-state}. This analysis has not
been performed before in the 
literature\footnote{The action $\Omega$ in the light-cone gauge
has been studied for boundary states in \cite{Bergman-Gaberdiel-open}, 
and for string states in \cite{Gimon}.}
and clarifies interesting subtleties about
the action of $\Omega$ on the boundary states. 
Adding up the information about the ${\cal I}_4$ projection on the
boundary states given in \cite{Gaberdiel-Stefanski}, we can deduce
the boundary states which will survive in the type I orbifold. The
next step is to put together these states to make up the D-brane
states of the theory. By the nature of the orbifold, we can deduce
that there are bulk and fractional BPS D-branes. Regarding the non-BPS
branes, some of them are truncated with twisted R-R charge,
similar to the type IIB orbifold. The rest of them
have no R-R charge at all, like in the type I theory before orbifolding. This latter 
type of brane can be seen as truncated branes of the type IIB
orbifold that get their R-R sector projected out by $\Omega$, 
or else, as non-BPS branes of type I that did not get too much affected by the
orbifold. 
On the other hand, those truncated branes of the type I orbifold 
that have twisted R-R charge 
can be seen either as truncated branes of the type IIB orbifold
that are not modified by the $\Omega$ projection, 
or as non-BPS branes of type I that receive a contribution 
from the twisted sector.

The paper is organised as follows. In section \ref{section2}
we present a thorough study of the action of $\Omega$ in the different
boundary state sectors. A summary of the results can be found in
Table \ref{table: state-invariance}. Using these results, 
in section \ref{section3} we give a classification of the BPS 
and non-BPS branes of the type I orbifold.
This classification is summarised in Tables
2 and 3.
Section \ref{section4} is concerned about the stability of the non-BPS branes
in the type I orbifold. In section \ref{section5} we discuss the 
conclusions of our analysis.
We include three appendices. Appendix \ref{appendixA}
explains in detail the action of the operator $\O$ on states in 
the asymmetric superghost picture $(-1/2,-3/2)$.
Appendix \ref{appendixA2} deals with the details of the action of
$\Omega$ on the twisted NS-NS sector.
Finally, Appendix \ref{appendixB} contains details about the
form of the crosscaps states used for the computations of this paper.

%%%%%%%%%%%%%%%%%%%%%%%%%%%%%%%%%%%%%%%%%%%%%%%%%%%%%%%%%%%%%%%%%%%%%%%%%
%%%%%%%%%%%%%%%%%%%%%%%%%%%%%%%%%%%%%%%%%%%%%%%%%%%%%%%%%%%%%%%%%%%%%%%%%
%%%%%%%%%%%%%%%%%%%%%%%%%%%%%%%%%%%%%%%%%%%%%%%%%%%%%%%%%%%%%%%%%%%%%%%%%

\section{Action of $\Omega$ on the Boundary States}
\label{section2}

%%%%%%%%%%%%%%%%%%%%%%%%%%%%%%%%%%%%%%%%%%%%%%%%%%%%%%%%%%%%%%%%%%%%%%%%%
%%%%%%%%%%%%%%%%%%%%%%%%%%%%%%%%%%%%%%%%%%%%%%%%%%%%%%%%%%%%%%%%%%%%%%%%%
%%%%%%%%%%%%%%%%%%%%%%%%%%%%%%%%%%%%%%%%%%%%%%%%%%%%%%%%%%%%%%%%%%%%%%%%%

We consider first the action of $\Omega$ on the open string
sectors. From the mode expansion of the fields we have for the oscillators 
\bea
\O \alpha_n^\mu \O^{-1} &=& \pm {\rm e}^{i n\pi} \alpha_n^\mu \, ,\nonumber \\
\O \psi_m^\mu \O^{-1} &=& \pm {\rm e}^{i m\pi} \psi_m^\mu \, ,
\eea
where the plus sign is for the NN directions and the minus sign is for the
DD directions. Moreover, $\O$ relates the DN and ND strings so there is no
definite action in these oscillators for these directions.
Furthermore, for the NS-vacuum in the $(-1)$ picture we have
\be
\Omega \ket{0}_{-1} = -i \ket{0}_{-1} \, ,
\ee
whereas the R-vacuum in the $(-1/2)$ picture transforms as:
\be
\Omega \ket{a}_{-1/2} = - \Gamma^{\nu_{p+1}} \cdots \Gamma^{\nu_{9-p}}
\ket{a}_{-1/2} \, ,
\ee
where $\nu_{p+1}, \dots \nu_{9-p}$ are the DD directions.

Regarding the oscillators of the
closed string, for convenience, 
we define $\Omega$ as
a combination of worldsheet parity-reversal
($\sigma$ $\rightarrow$ $2\pi - \sigma$)
and the GSO-projection. With this definition it has the same action 
on the physical states as worldsheet parity-reversal only.
On the oscillators it acts by simply exchanging left and right sectors:
\be
\O \alpha_n \O^{-1} = \tilde{\alpha_n} \, ,\qquad
\O \psi_n \O^{-1} = \tilde{\psi_n} \, ,
\ee
and analogously for the right sector. 

Similarly, $\O$ exchanges the
left and right sectors of the (super)ghosts:
\bea
\O b_n \O^{-1} = {\tilde b}_n \, &,& \qquad \O c_n \O^{-1} = {\tilde c}_n
\, ,\nonumber \\
\O \beta_m \O^{-1} = {\tilde \beta}_m \, &,& \qquad 
\O \gamma_m \O^{-1} = {\tilde \gamma}_m \, ,
\eea
and similarly for the right sector.
The action of $\Omega$ on the NS-NS and R-R ground states for
untwisted and twisted sectors are studied below, and the results are
given in equations (\ref{omega-groundstate-1}), 
(\ref{omega-groundstate-2}), (\ref{(2.34)}) and
(\ref{omega-groundstate-3})

\subsection{The Untwisted NS-NS Sector}
 
The untwisted NS-NS boundary state\footnote{We use the conventions of
\cite{Divecchia-Liccardo} for the definition of the boundary states.
See also \cite{particle} for details.} 
\be
\ket{D(r,s),\eta}_{\rm NS,U} =
{T_{(r,s)} \over 2} \, \ket{D(r,s)_X}
\ket{D(r,s)_\psi,\eta}_{\rm NS,U} \ket{D(r,s)_{gh}} 
\ket{D(r,s)_{sgh},\eta}_{\rm NS} \, ,
\ee
is constructed upon the
NS-NS ground state in the $(-1,-1)$ picture. The action of $\O$
on this ground state is given by:
\be
\O \left( \ket{0}_{-1} \otimes \widetilde{\ket{0}}_{-1} 
\right) = \widetilde{\ket{0}}_{-1} \otimes {\ket{0}}_{-1}
=  - \ket{0}_{-1} \otimes \widetilde{\ket{0}}_{-1} \, .
\label{omega-groundstate-1}
\ee
On the other hand, since $\O$ exchanges left and right oscillators,
the overall effect is a change from $\eta$ to  $-\eta$,
hence we find:
\be
\O \ket{D(r,s),\eta}_{\rm NS,U} = - \ket{D(r,s),- \eta}_{\rm NS,U} \, .
\ee
Therefore, the
GSO projected untwisted NS-NS boundary state
\be
\ket{D(r,s)}_{\rm NS,U} = 
{1 \over 2}\left(\ket{D(r,s), +}_{\rm NS,U} - 
\ket{D(r,s),-}_{\rm NS,U}\right)
\, ,
\ee
is invariant under $\O$, for any $p=r+s$.
        
%%%%%%%%%%%%%%%%%%%%%%%%%%%%%%%%%%%%%%%%%%%%%%%%%%%%%%%%%%%%

\subsection{The Untwisted R-R Sector}

Consider now the untwisted R-R part.
The R-R ground state in the $(-1/2,-1/2)$ picture
has the following transformation property under $\O$:
\be
\O \left( \ket{A}_{-1/2} \otimes \widetilde{\ket{B}}_{-1/2} 
\right) = \widetilde{\ket{A}}_{-1/2} \otimes {\ket{B}}_{-1/2}
=  - \ket{B}_{-1/2} \otimes \widetilde{\ket{A}}_{-1/2} \, .
\ee
On the other hand, 
the R-R boundary state in the covariant formalism
is most conveniently written in the asymmetric 
superghost picture $(-1/2,-3/2)$ \cite{Divecchia-etal}:
\be
\ket{D(r,s),\eta}_{\rm R,U} =
{Q_{(r,s)} \over 2} \, \ket{D(r,s)_X}
\ket{D(r,s)_\psi,\eta}_{\rm R,U} \ket{D(r,s)_{gh}} 
\ket{D(r,s)_{sgh},\eta}_{\rm R} \, ,
\ee
where $r+s=p$ odd in type IIB.
The fermionic matter and the superghost components have zero-mode 
contributions
\be 
\ket{D(r,s),\eta}^{(0)}_{\rm R,U} = \ket{D(r,s)_{\psi}, \eta}^{(0)}_{\rm R,U}
\ket{D(r,s)_{sgh}, \eta}^{(0)}_{\rm R} \, ,
\ee
given by
\bea
\ket{D(r,s),\eta}^{(0)}_{\rm R,U} &=&
{\rm e}^{i\eta \gamma_0 \tilde{\beta}_0} 
{\cal M}_{AB} \, \ket{A}_{-1/2} \otimes \widetilde{\ket{B}}_{-3/2} \, ,\\
{\cal M}_{AB}(\eta) &=& 
\left( {\cal C}_{(10)} 
\Gamma^0 \cdots \Gamma^p {1 + i\eta \Gamma_{11} \over
1+ i\eta}\right)_{AB} \, ,\qquad p=r+s 
\, ,\nonumber 
\eea
where $A,B$ are spinor indices of $SO(1,9)$ and ${\cal C}_{(10)}$
is the charge conjugation matrix in the corresponding representation
of the $10$-dimensional $\Gamma$-matrices.

Note that under the action of $\Omega$, the superghost pictures
seem to be swapped between left and right:
\be
\O \left( \ket{A}_{-1/2} \otimes \widetilde{\ket{B}}_{-3/2} 
\right) = \widetilde{\ket{A}}_{-1/2} \otimes {\ket{B}}_{-3/2}
=  - \ket{B}_{-3/2} \otimes \widetilde{\ket{A}}_{-1/2} \, .
\ee
The commutation of the superghosts in order to recover 
the $(-1/2,-3/2)$ could yield a
phase that we must calculate.
In Appendix \ref{appendixA} we give
the details about how to obtain the action of $\Omega$ in the asymmetric 
picture. Here we present directly the result for the superghost sector:
\be
\Omega \ket{ D(r,s)_{sgh}, \eta}_{\rm R} = 
-i \eta \ket{D(r,s)_{sgh}, -\eta}_{\rm R}
\, .
\label{The-result}
\ee
Regarding the fermion ground state we have
\be
\O \left( \ket{A} \otimes \widetilde{\ket{B}} 
\right) =  - \ket{B} \otimes \widetilde{\ket{A}} \, ,
\label{omega-groundstate-2}
\ee
hence the effect of $\Omega$ on the matrix ${\cal M}_{AB}$ is a transposition:
\bea
{\cal M}(\eta)^T &=&  (-1)^{p+ {1 \over 2}p (p+1)}
\left( {\cal C} \Gamma^0 \cdots \Gamma^p {1 + (-1)^p i \eta \Gamma_{11} \over
1+ i\eta}\right) \nonumber \\
&=& - i\eta (-1)^{p+ {1 \over 2}p(p+1)}\, {\cal M} (-\eta)\, .
\nonumber
\eea
We then find 
\be
\Omega \ket{D(r,s)_{\psi}, \eta}^{(0)}_{\rm R}
= - i\eta (-1)^{ {1 \over 2}p(p+1)} \, \ket{D(r,s)_{\psi}, -\eta}^{(0)}_{\rm R}
\, .
\ee
In the oscillator part, $\Omega$ exchanges tilded oscillators
with untilded ones, yielding an overall change of $\eta$ to $-\eta$.
Using this fact and the result (\ref{The-result})
we finally find:
\be
\Omega \ket{ D(r,s)_{\psi}, \eta}_{\rm R} \ket{ D(r,s)_{sgh}, \eta}_{\rm R}
= - (-1)^{{1 \over 2} p(p+1)} \, 
\ket{D(r,s)_{\psi}, -\eta}_{\rm R} \ket{ D(r,s)_{sgh}, -\eta}_{\rm R}
\, ,
\ee
so that the GSO-invariant R-R boundary state
\be
\ket{D(r,s)}_{\rm R,U} = 
{1 \over 2}\left(\ket{D(r,s), +}_{\rm R,U} + \ket{D(r,s),-}_{\rm R,U}\right)
\, ,
\ee
is $\Omega$-invariant for $p=1,5,9$, as expected for the physical BPS
D-branes in the type I theory in the critical dimension.

%%%%%%%%%%%%%%%%%%%%%%%%%%%%%%%%%%%%%%%%%%%%%%%%%%%%%%%%%%%%%%%%%%%%%%%%%%

Likewise, we can determine the $\O$-projection on the D-branes by using the
form of the quantum R-R string state in the $(-1/2,-3/2)$ 
picture\footnote{This particular approach presents the advantage over
the usual one in the $(-1/2,-1/2)$ picture of making possible
the description of spacetime-filling D-branes.}, given
in terms of the R-R gauge potential:
\bea
\ket{C^{(p+1)}} &=& {1 \over 2 (p+1)! \sqrt{2}} 
\, C^{(p+1)}_{\mu_1 \dots \mu_{p+1}}
\left( \left( {\cal C}_{(10)}\Gamma^{\mu_1 \dots \mu_{p+1}} \Pi_+ \right)_{AB}
\, {\rm cos} \gamma_0 \tilde{\beta}_0 \right. \\
&&\hspace{3cm} + \,
\left. \left( {\cal C}_{(10)} \Gamma^{\mu_1 \dots \mu_{p+1}} \Pi_- \right)_{AB}
\, {\rm sin} \gamma_0 \tilde{\beta}_0 \right)
\, \ket{A}_{-1/2} \otimes \widetilde{\ket{B}}_{-3/2} \, .\nonumber
\eea
We refer again to Appendix \ref{appendixA} for the details 
of this derivation. We state here the result:
\be
\Omega \, \ket{C^{(p+1)}} = 
- (-1)^{{1 \over 2}p(p+1)} \, \ket{C^{(p+1)}} \, .
\ee
Thus it is invariant for $p=1,5,9$;
the same values of $p$ for which the R-R 
boundary states are $\Omega$-invariant.

%%%%%%%%%%%%%%%%%%%%%%%%%%%%%%%%%%%%%%%%%%%%%%%%%%%%%%%%%%%%

\subsection{The Twisted R-R Sector}

The orbifold of type IIB on $T^4/{\cal I}_4$
has 16 6-dimensional fixed planes with $(2,0)$ supersymmetry each.
The R-R twisted ground state consists of two spinors
of $SO(1,5)$ of the same chirality, either ${\bf 4}$ or ${\bf 4}^\prime$.
Therefore, the twisted R-R sector can contain the following field strengths:
\be
{\bf 4} \otimes {\bf 4} = [1] + [3]_+ \, ,\qquad
{\bf 4}^\prime \otimes {\bf 4}^\prime = [1] + [3]_- \, .
\label{product-decomposition}
\ee
Depending on which chirality we choose for the spinors, the 3-form
field strength will be self-dual or anti self-dual.
Accordingly, we have twisted R-R boundary states
$\ket{D(r,s)}_{\rm R,T}$ for $r=-1,1,3$. As we will see below, 
only those states with $r=-1,3$ will survive the $\O$ projection,
that is, only the twisted R-R scalar potential
in (\ref{product-decomposition}) survives \cite{Bianchi-2,Gimon}.
However, note that (\ref{product-decomposition}) does not directly show the
existence of a twisted 6-form potential, which was assumed in \cite{Gimon}
and would be responsible for the twisted R-R tadpoles.
This potential would be needed as well in order to account for an $r=5$ 
twisted R-R boundary state that one can construct, as we show below.
We see by direct construction of the corresponding
quantum state that such a R-R potential does exist but it
does not survive the $\O$ projection\footnote{On the other hand, 
one could choose the projection 
$\Omega= -1$ in the twisted sector, defined as
the $\Omega \, {\cal J}$ projection in \cite{Polchinski-K3}. 
In that case the twisted R-R 6-form 
potential would survive in the orientifold.}.

Consider first the action of $\Omega$ on the boundary states.
The twisted R-R boundary state is given by
\be
\ket{D(r,s),\eta}_{\rm R,T} =
{{\tilde Q}_{(r,s)} \over 2} \, \ket{D(r,s)_X}_{\rm T}
\ket{D(r,s)_\psi,\eta}_{\rm R,T} \ket{D(r,s)_{gh}} 
\ket{D(r,s)_{sgh},\eta}_{\rm R} \, ,
\ee
which is constructed upon the twisted R-R ground state
in the asymmetric $(-1/2,-3/2)$ picture \cite{particle}.
This sector contains zero-modes:
\be
\ket{D(r,s),\eta}^{(0)}_{\rm R,T} = 
{\rm e}^{i\eta \gamma_0 \tilde{\beta}_0} 
{M}_{ab} \, \ket{a}_{-1/2,{\rm T}} 
\otimes \widetilde{\ket{b}}_{-3/2,{\rm T}} \, ,
\ee
with
\be
{M}_{ab}(\eta)  = \left( {\cal C}_{(6)} 
\gamma^0 \cdots \gamma^r {1 + i\eta \gamma \over
1+ i\eta}\right)_{ab} \, ,
\ee
where $a,b$ are spinor indices of $SO(1,6)$ and
${\cal C}_{(6)}$ is the charge conjugation matrix associated
to the $SO(1,5)$ $\gamma$-matrices, for which we use the same
conventions as in \cite{particle}.

The action of $\O$ on this ground state 
results as before an exchange of pictures between left and right,
%\be
%\O \left( \ket{a}_{-1/2, {\rm T}} \otimes \widetilde{\ket{b}}_{-3/2,{\rm T}} 
%\right) = \widetilde{\ket{a}}_{-1/2,{\rm T}} 
%\otimes {\ket{b}}_{-3/2,{\rm T}}
%=  
%- \ket{b}_{-3/2,{\rm T}} \otimes \widetilde{\ket{a}}_{-1/2,{\rm T}} \, .
%\label{t-groundstate-twist}
%\ee
hence we could proceed as before.
Since the superghost part is the same as in the untwisted sector,
the action of $\O$ on the  $(-1/2,-3/2)$ picture 
is the same as given before in 
(\ref{The-result}), and the effect on the zero-mode matrix
$M_{ab}$ is again a transposition:
\bea
M(\eta)^T  &=& - (-1)^{r+ {1 \over 2}r (r+1)}
\left( {\cal C}_{(6)} \gamma^0 \cdots \gamma^r {1 + (-1)^r i \eta \gamma \over
1+ i\eta}\right) \nonumber \\
&=& - i\eta (-1)^{{1 \over 2}r(r+1)}\, M (-\eta)\, ,
\nonumber
\eea
where we have used the fact that fractional and truncated branes
in type IIB on $T^4/{\cal I}_4$ only exist for $r$ odd.
Finally, using the above results and the fact
that the twisted spinor indices 
anti-commute \cite{Polchinski-K3,orbifold-vertices}, we find:
\be
\O \ket{D(r,s),\eta}_{\rm R,T}^{(0)} =
(-1)^{{1 \over 2}r (r+1)} \, \ket{D(r,s),- \eta}_{\rm R,T}^{(0)} \, ,
\ee
hence the GSO-invariant twisted R-R boundary state
\be
\ket{D(r,s)}_{\rm R,T} = 
{1 \over 2}\left(\ket{D(r,s), +}_{\rm R,T} + \ket{D(r,s),-}_{\rm R,T}\right)
\, ,
\ee
is $\Omega$ invariant for $r=-1,3$. For fractional branes $s$ is even and
for truncated
branes $s$ is odd \cite{Gaberdiel-Stefanski}. Thus we find that
there are non-BPS (truncated) branes 
$\ket{D(-1,s)}$ and $\ket{D(3,s)}$ branes,
for $s$ odd, in type I on $T^4/{\cal I}_4$.
For the fractional (BPS) branes we must still find out 
which NS-NS twisted boundary states survive the projection. This is
done in the next subsection.

We can find the same result using the quantum states of the R-R twisted
sector in the $(-1/2,-3/2)$ picture:
\bea
\ket{C^{(r+1)}}_{\rm T} &=& 
{1 \over \sqrt{2} (p+1)!} C^{(r+1)}_{\mu_1 \dots \mu_{r+1}} \, 
\left( \left( {\cal C}_{(6)}\gamma^{\mu_1 \dots \mu_{p+1}} \Pi_+ \right)_{ab}
\, {\rm cos} \gamma_0 \tilde{\beta}_0 \right.\\
&&\hspace{3cm}+ \,
\left. \left( {\cal C}_{(6)}\gamma^{\mu_1 \dots \mu_{p+1}} \Pi_- \right)_{ab}
\, {\rm sin} \gamma_0 \tilde{\beta}_0 \right)
\, \ket{a}_{-1/2,{\rm T}} \otimes \widetilde{\ket{b}}_{-3/2,{\rm T}} \, .
\nonumber
\eea
Using (\ref{cos-sin-twist}) from Appendix \ref{appendixA} we find
\be
\O \, \ket{C^{(r+1)}}_{\rm T} =
(-1)^{{1 \over 2}r(r+1)} \, \ket{C^{(r+1)}}_{\rm T} \, ,
\ee
hence the twisted scalar and its dual, the 4 form potential, survive 
the $\O$ projection, whereas the 6-form potential
does not survive, which is in agreement with the results of 
\cite{Bianchi-1,Bianchi-2,Gimon}.

%%%%%%%%%%%%%%%%%%%%%%%%%%%%%%%%%%%%%%%%%%%%%%%%%%%%%%%%%%%%

\subsection{The Twisted NS-NS Sector}

The boundary state in the twisted  NS-NS sector
\be 
\ket{D(r,s),\eta}_{\rm NS,T} =
{\tilde{T}_{(r,s)} \over 2} \, \ket{D(r,s)_X}_{\rm T}
\ket{D(r,s)_\psi,\eta}_{\rm NS,T} \ket{D(r,s)_{gh}} 
\ket{D(r,s)_{sgh},\eta}_{\rm NS} 
\, 
\ee
is constructed upon the NS-NS ground state in
the $(-1,-1)$ superghost picture.
This boundary state contains a sector for the fermion zero-modes coming
from the compact directions $6$, $7$, $8$ and $9$ :
\be
\ket{D(r,s),\eta}^{(0)}_{\rm NS,T} =
{m}_{\alpha\beta}(\eta)  \, \ket{\alpha}_{-1,{\rm T}} 
\otimes \widetilde{\ket{\beta}}_{-1,{\rm T}} \, ,
\ee
where $\alpha,\beta$ are spinor indices 
of $SO(4)$. Furthermore, from the overlapping condition for the zero
modes, we find:
\be
{m}_{\alpha\beta} (\eta)  = 
\left( {\cal C}_{(4)} \Pi_{(s)}  { 1 - i\eta \bar{\gamma} \over 1 -i \eta} 
\right)_{\alpha\beta}
\, ,
\ee
where $\Pi_{(s)}$ is the product of $s$ gamma-matrices:
\bea
\Pi_{(s)} &=& \bar{\gamma}^6 \bar{\gamma}^7 \cdot \dots \, ,\qquad s \neq 0 \, ,\nonumber \\
\Pi_{(s)} &=& 1  \, ,\qquad s = 0 \, ,
\eea
and $\bar{\gamma}= -  
\bar{\gamma}^6 \bar{\gamma}^7 \bar{\gamma}^8 \bar{\gamma}^9$ is the
chirality matrix in four Euclidean dimensions.
Moreover, ${\cal C}_{(4)}$ represents the charge-conjugation matrix
in the present context. Our conventions for
the $SO(4)$ gamma-matrices and the action
of the fermionic zero-modes on the ground state 
can be found in Appendix \ref{appendixA2}.

The ground state, which is constructed with
anti-commuting spin fields of the same chirality, 
transforms as follows under $\O$:
\be
\Omega \left( \ket{\alpha}_{\rm -1,T} \otimes \widetilde{\ket{\beta}}_{\rm
-1,T}\right)
= ({\bar{\gamma}})^{\alpha}{}_{\delta} \,
\ket{\beta}_{\rm -1,T} \otimes \widetilde{\ket{\delta}}_{\rm -1,T}
= ({\bar{\gamma}})^{\beta}{}_{\delta} \, 
{\ket{\delta}}_{\rm -1,T}  \otimes \widetilde{\ket{\alpha}}_{\rm -1,T}
\, .
\label{(2.34)}
\ee
A proof of this result is presented in Appendix \ref{appendixA2}.
Using this result we obtain the following action of $\Omega$ on the
boundary state corresponding to the vacuum sector:
\bea
\Omega \, \ket{D(r,s),\eta}^{(0)}_{\rm NS,T}
&=&
\left( m(\eta)^T \, \bar{\gamma} \right)_{\alpha \beta} 
\ket{\alpha}_{\rm -1,T} \otimes \widetilde{\ket{\beta}}_{\rm -1,T}
\nonumber \\
&=& - (-1)^{ {1 \over 2} s (s-1)} \,
\ket{D(r,s),- \eta}^{(0)}_{\rm NS,T}
\, .
\eea
with $s$ even.
Note that the twisted NS-NS boundary state is ${\cal I}_4$-invariant 
for $s=even$ \cite{Gaberdiel-Stefanski}.
On the other hand, in the sector bilinear in the oscillators $\O$ acts again
by changing $\eta$ by $-\eta$.
Thus we finally deduce:
\be
\Omega \, \ket{D(r,s), \eta}_{\rm NS,T} = - (-1)^{ {1 \over 2} s (s-1)}
\, \ket{D(r,s), - \eta}_{\rm NS,T} \, ,
\ee
hence the GSO-projected boundary state
\be
\ket{D(r,s)}_{\rm NS,T} = 
{1 \over 2}\left( \ket{D(r,s),+}_{\rm NS,T} +
\ket{D(r,s),-}_{\rm NS,T} \right) 
\, ,\qquad s\, = \, even
\ee
is $\Omega$-invariant only for $s=2$.
Accordingly, only branes with $s=2$ in type I on $T^4/{\cal I}_4$
carry a NS-NS twisted sector.

%\section{Non-BPS Crosscaps}
%In order to have open-closed string consistency,
%the open strings in the amplitude between two non-BPS branes
%after the orientifold projection should have also the correct
%projection under $\Omega$. Namely
%\be
%{1 \over 4} (1 - (-1)^{F+G} )(1 + \Omega) \, .
%\ee
%this requires to include a crosscap that accounts for the
%cylinder with the insertion of $\Omega$. 
%For non-BPS branes this would correspond to some sort of non-BPS crosscaps.
%After a conversation with Gaberdiel, it turned out that he have also 
%considered
%that possibility and that it appears naturally in the
%asymmetric orbifold that he has constructed together with
%C.Angelantonj and R.Blumenhagen in this work in progress I mentioned to you, 
%about cancelling tadpoles with non-BPS branes.
%He told me that these non-BPS crosscaps
%have only sectors in the NS-NS,T and R-R,T.
%It would be interesting if we can also understand them in the present context,
%in order to make the open string spectrum in the non-BPS D-branes
%consistent with the orientifold.

%%%%%%%%%%%%%%%%%%%%%%%%%%%%%%%%%%%%%%%%%%%%%%%%%%%%%%%%%%%%
%%%%%%%%%%%%%%%%%%%%%%%%%%%%%%%%%%%%%%%%%%%%%%%%%%%%%%%%%%%%
%%%%%%%%%%%%%%%%%%%%%%%%%%%%%%%%%%%%%%%%%%%%%%%%%%%%%%%%%%%%

\section{Non-BPS Branes in Type I on $T^4/\Z_2$}
\label{section3}

%%%%%%%%%%%%%%%%%%%%%%%%%%%%%%%%%%%%%%%%%%%%%%%%%%%%%%%%%%%%
%%%%%%%%%%%%%%%%%%%%%%%%%%%%%%%%%%%%%%%%%%%%%%%%%%%%%%%%%%%%
%%%%%%%%%%%%%%%%%%%%%%%%%%%%%%%%%%%%%%%%%%%%%%%%%%%%%%%%%%%%

Before describing in detail the stability of the non-BPS branes in 
the type I orbifold, we summarise the brane spectrum that we have found
with the analysis based on the boundary states.
Recall that the D-branes are denoted by $\ket{D(r,s)}$,
with $r$ the number of Neumann directions along the orbifold fixed plane and
$s$ the number of Neumann directions along $T^4$. From 
the point of view of the closed string the
orientifold is described in terms of crosscaps. Since there are orientifold
9- and 5- planes, we must consider both 9-crosscap
$\ket{C9}$ and 5-crosscaps $\ket{C5}$.
The type I orbifold can be considered as an orientifold
of type IIB on $T^4/{\cal I}_4$. Accordingly the boundary states for 
the branes in the type I orbifold theory
will be modified by the addition of the crosscaps:
\be
\ket{D} \longrightarrow \ket{D}_{T^4/\Z_2} =
{1 \over \sqrt{2}} \left( \ket{D} + \ket{C9} + \ket{C5} \right) \, ,
\ee
where we would introduce one 5-crosscap for each fixed plane,
and we have chosen this particular 
normalisation to account for the normalisation of 
the orientifold projectors in the open string channel.

\TABLE[ht!]{
\renewcommand{\arraystretch}{1.5}  
%\begin{center}
\begin{tabular}{|c|c|c|}
\hline
Boundary State 	& $\O$-invariance 	& ${\cal I}_4$-invariance \\
\hline\hline
NS-NS untwisted & $\forall$  $r,s$ 	&  $\forall$  $r,s$ \\
\hline
R-R untwisted 	& $r+s = 1,5,9$ 	& $s=$ even \\
\hline
R-R twisted   	& $r=-1,3$              & $\forall$  $r,s$ \\
\hline
NS-NS twisted   & $s=2$		& $s=$ even	\\
\hline
\end{tabular}
%\end{center}  
\caption{\label{table: state-invariance} 
\small {\bf $\O$ and ${\cal I}_4$ -invariance of the boundary states.} 
This table shows for which values of $r$ and $s$ each boundary state is 
invariant under the operations $\Omega$ and ${\cal I}_4$.}
\renewcommand{\arraystretch}{1}  
}

In type I on $T^4/\Z_2$ we find also BPS fractional branes,
which have states in all sectors twisted and 
untwisted, and include the crosscaps as well:
\be
\ket{D(r,s)}_{f,\,  T^4/\Z_2} =
{1 \over \sqrt{2}}
\left( \ket{D(r,s)}_f + \ket{C9} + \ket{C5} \right) \, .
\ee
The values of $r$ and $s$ for which they exist are such that
the boundary state in every sector is invariant under
both $\Omega$ and ${\cal I}_4$ besides being GSO-invariant. From the results
of the previous section,  shown in table \ref{table: state-invariance},
we find these values to be $r=-1,3$ and $s=2$,  i.e. the orbifold theory
has an {\it instantonic} D$1$ and a D$5$ as fractional branes. 
%This is consistent since the D$9$-brane, 
%being a space-time filling brane, cannot be localised at the orbifold
%fixed points.

\TABLE[p]{
\renewcommand{\arraystretch}{1.5}  
%\begin{center}
\begin{tabular}{|c|c|}
\hline
BPS branes& \\
\hline\hline
Fractional \& Bulk  &  $r=-1,3$ \\
branes	          &  $s=2$   \\
\hline
Bulk  branes      & $r=1,5$ \\
	          & $s=0,4$ \\
\hline
\end{tabular}
%\end{center}  
\caption{\label{table: BPS-branes}
\small {\bf BPS branes in type I on $T^4/\Z_2$.} 
This table shows for which $(r,s)$ we can have either
BPS fractional or bulk branes in the type I orbifold.}
\renewcommand{\arraystretch}{1}  
}

BPS Bulk branes can be described as branes with only untwisted sectors.
A possibility is to have $r$ and $s$ such that the untwisted sectors
are invariant, but not the twisted sectors. This yields
$r=1,5$ and $s=0,4$, which accounts for 
$D1$, $D5$ and $D9$ branes. On the other hand,
two fractional branes with opposite twisted R-R charge
can join and produce a bulk brane that can move off the
fixed planes. Therefore we also can have bulk branes
for the same values of $r$ and $s$ as for fractional branes.
Bulk branes can be represented as follows: 
\be
\ket{D(r,s)}_{b,\,  T^4/\Z_2} =
{1 \over \sqrt{2}}
\left( \ket{D(r,s)}_b + \ket{C9} + \ket{C5} \right) \, .
\ee

\TABLE[p]{
\renewcommand{\arraystretch}{1.5}  
%\begin{center}
\begin{tabular}{|c|c|c|}
\hline
Non-BPS branes& 	& \\
\hline\hline
$\Z$-charge &  $r=-1,3$    &  \\
	    &  $s=0,1,3,4$ &  \\
\hline
$\Z_2$-charge &  $r=1,5$  &  $r=$ even \\
	    &  $s=0$  &  $s=0$ \\
\hline
\end{tabular}
%\end{center}  
\caption{\label{table: brane-scan}
\small {\bf Non-BPS branes in type I on $T^4/\Z_2$.} 
Non-BPS branes can have $\Z$-charge, if they have a twisted R-R sector,
or $\Z_2$ charge if they do not have any R-R sector at all.
The table shows for which $(r,s)$ the boundary states exist
in the type I orbifold.}
\renewcommand{\arraystretch}{1}  
}

In the type I orbifold, only those truncated branes whose twisted R-R
state survives the orbifold symmetry are $\Z$-charged non-BPS branes.
Moreover, some non-BPS branes of type I have values
$r$ and $s$ such that the twisted R-R charge in the type I 
orbifold exist. Thus these branes are also truncated in the orbifold.
We find that $\Z$-charged non-BPS branes can exist 
for $r=-1,3$ and $s=0,1,3,4$, since for these values 
$\ket{D(r,s)}_{\rm R,T}$ is $\Omega$-invariant but
$\ket{D(r,s)}_{\rm R}$ is not. 
These branes can be represented by:
\be
\ket{D(r,s)}_{t, T^4/\Z_2} = {1 \over \sqrt{2}}
\left( \ket{D(r,s)}_{t} + \ket{C9} + \ket{C5} 
\right) \, .
\ee
\noindent These branes include the non-BPS D-instanton and D$7$ brane
of type I, which have twisted R-R charge in the type I orbifold.
This $\Z$-charged D$7$ brane appears only for the orientation $(3,4)$.

The last possibility consists of those branes which
have neither untwisted nor twisted R-R charge.
These include those branes which are BPS in the type IIB orbifold, 
but become non-BPS in the orientifold, and also
those branes in type I which do not obtain any R-R charge
after the orbifolding. They are described solely by
an untwisted NS-NS sector, and in the type I orbifold can be 
represented by:
\be
\ket{D(r,s)}_{t,T^4/\Z_2} = 
{1 \over \sqrt{2}} \left( 
\ket{D(r,s)}_{\rm NS,U} + \ket{C9} + \ket{C5} \right) \, .
\ee
In principle, the values of $r$ and $s$ for which they can exist are
those for which no R-R sector survives.
The two possibilities would be
$r=1,5$ and $s=1,2,3$, or $r=$even (including $r=0$) and $s=0,...,4$.
These branes would include the non-BPS D$0$, 
as well the D$7$ and D$8$ branes of type I, which
can appear with different orientations, namely,
$(5,2)$, $(4,3)$, $(5,3)$, $(4,4)$.
However, as will be explained in the next section, only $(r,0)$ branes
can be properly defined as $\Z_2$-charged branes.
We will also prove in the next
section that only the $(0,0)$ and the $(-1,0)$ branes are
in fact stabilised by the action of the orientifold.

%%%%%%%%%%%%%%%%%%%%%%%%%%%%%%%%%%%%%%%%%%%%%%%%%%%%%%%%%%%%%%%%%%
%%%%%%%%%%%%%%%%%%%%%%%%%%%%%%%%%%%%%%%%%%%%%%%%%%%%%%%%%%%%%%%%%%
%%%%%%%%%%%%%%%%%%%%%%%%%%%%%%%%%%%%%%%%%%%%%%%%%%%%%%%%%%%%%%%%%%

\section{Stability of the Non-BPS Branes}
\label{section4}

%%%%%%%%%%%%%%%%%%%%%%%%%%%%%%%%%%%%%%%%%%%%%%%%%%%%%%%%%%%%%%%%%%
%%%%%%%%%%%%%%%%%%%%%%%%%%%%%%%%%%%%%%%%%%%%%%%%%%%%%%%%%%%%%%%%%%
%%%%%%%%%%%%%%%%%%%%%%%%%%%%%%%%%%%%%%%%%%%%%%%%%%%%%%%%%%%%%%%%%%

In this section we study the stability of the non-BPS branes
presented in the previous section. For this analysis 
it is crucial to notice that
the tadpole-cancelling D9 and D5 branes in type I on
$T^4/{\cal I}_4$ are bulk 
branes\footnote{This can be deduced from the fact that the twisted
R-R state associated with the 6-form potential is odd under $\Omega$, which
agrees with the fact that the tadpole associated
with the twisted 6-form is identically zero
\cite{Bianchi-1,Bianchi-2,Gimon}.}
The normalisation of the boundary states in the orbifold
involves the trace of the representation matrices of the orientifold
group on the open string sector. This trace is exactly zero
for the case of the twisted R-R 6-form, hence the D9 and D5 branes
do not carry a twisted R-R sector, and by supersymmetry they do not carry
any twisted NS-NS sector either.
Therefore the 9- and 5- crosscaps of the type I orbifold
have boundary states in the untwisted sectors 
only\footnote{Consistency conditions of type I on the 
orbifold $T^4/{\cal I}_4$ using the
boundary state formalism was first considered in \cite{Ishibashi}.}:
\be
\ket{C9}=\ket{C9}_{\rm NS,U} + \ket{C9}_{\rm R,U} \, ,\qquad
\ket{C5}=\ket{C5}_{\rm NS,U} + \ket{C5}_{\rm R,U} \, .
\ee
The detailed form of these crosscaps is given in Appendix
\ref{appendixB}.
We start by carrying out first in detail the analysis of the stability for 
the $\Z_2$-charged D-particle, which will illustrate the more general case
for $(r,s)$ branes considered next. We then finish by considering the
$\Z$-charged branes.

%%%%%%%%%%%%%%%%%%%%%%%%%%%%%%%%%%%%%%%%%%%%%%%%%%%%%%%%%%%%%%%%%%

\subsection{The Stable Non-BPS D-particle with $\Z_2$ Charge}

In this subsection we explicitly carry out the computations for the D-particle and
establish the formalism to show various consistencies. In the next subsection, this
analysis can be simply used to obtain the required results for $(r,s)$ branes
without repeating the technical details. 
The boundary state for the D-particle in
the orbifold theory 
is given by: 
\be \ket{D0}_{T^4/\Z_2} =
{1 \over \sqrt{2}} \left( \ket{D0}_{\rm NS,U} 
+ \ket{C9} + \ket{C5} \right) \, ,
\ee
where $\ket{D0}_{\rm NS,U}$ denotes 
the GSO and ${\cal I}_4$ invariant NS-NS boundary state written
as a linear combination of the D-particle and its image in type IIB on $T^4/{\cal
I}_4$. Thus we have
\bea
\ket{D0}_{\rm NS,U} &=& 
{1 \over 2\sqrt{2}}\left( \ket{D0,a^\alpha,y^i,+}_{\rm NS,U} +
\ket{D0,a^\alpha,-y^i,+}_{\rm NS,U} \right. \\ 
&&\hspace{2cm} \left.- \ket{D0,a^\alpha,y^i,-}_{\rm NS,U}  -
\ket{D0,a^\alpha,-y^i,-}_{\rm NS,U}  \right) \, ,\nonumber
\eea
with $a^\alpha$ and $y^i$ the positions of the D-particle in the non-compact
and compact directions, respectively. For these branes, it is understood that $y^i$
is not the origin. The crosscap states are similarly taken in GSO
invariant combinations. 

We want to test whether there are any tachyons in the open string spectrum.
The relevant open string amplitudes to be considered are given by
the annulus and M${\ddot {\rm o}}$bius amplitudes, which can be
obtained from the following closed string amplitudes:
\bea
{\cal A} &=& {1 \over 2} {}_{\rm NS,U}\bra{D0} {\cal D} 
\ket{D0}_{\rm NS,U} \, ,\\
{\cal M}_9 + {\cal M}_5 &=& {1 \over 2} {}_{\rm NS,U}\bra{D0} {\cal D} 
\ket{C9} + 
{1 \over 2} {}_{\rm NS,U}\bra{D0} {\cal D} 
\ket{C5} + c.c \, ,\nonumber
\eea
where ${\cal D}$ is the closed string propagator
\be
{\cal D} = {\alpha^\prime \over 4\pi}
\int_{|z| \leq 1} {d^2 z \over |z|^2} z^{L_0 -a}
{\bar z}^{{\tilde L}_0 -a} \, ,
\ee
where $a=1/2$ for the untwisted NS-NS sector and
$a=0$ in the R-R sector (twisted and untwisted) and the
twisted NS-NS sector.
The crosscaps will interact with the $\Z_2$-charged D-particle
only through the untwisted NS-NS sector.

Note that the normalisation of the brane is  modified by the fact that
the orbifold is compact, namely:
\be
\ket{D0,\eta}_{\rm NS,U} = N_0 \,
\ket{D0_X}\ket{D0_\psi,\eta}_{\rm NS,U}\ket{D0_{gh}}
\ket{D0_{sgh}, \eta}_{\rm NS} \, ,
\ee
with
\be
N_0 = {\sqrt{2}}{T_0 \over 2} \left({2\pi R \over \Phi} \right)^2
(2\pi R)^{-4} \, ,
\ee
where the radius of each of the four compact coordinates is taken to be $R$ and 
$T_0$ is the tension of the BPS D-particle in type IIA theory:
\be
T_0 = \sqrt{\pi} (2\pi \sqrt{\alpha^\prime})^{3} \, .
\ee
We have introduced the self-dual volume $\Phi$ defined 
as\footnote{See \cite{Divecchia-Liccardo} for more details.}
\be
\bra{n,m} n^\prime m^\prime \rangle = \Phi \, \delta_{n n^\prime}
\delta_{m m^\prime} \, ,
\ee
for each compact direction. 
Moreover, the bosonic sector is modified as follows:
\be
\ket{D0_X} = \delta^{(5)}(q^\alpha - a^\alpha)
\,
\left(\sum\limits_{n \in \Z} {\rm e}^{i q_n {n \over R}}
\right)^{4}
\, \times \, \prod\limits_{n=1}^{\infty}
{\rm e}^{-{1 \over n} \alpha_{-n} \cdot {\cal S} \cdot
\tilde{\alpha}_{-n}}
\, \ket{k,m,n=0} \, .
\ee
With this information and using the 
explicit form of the crosscaps given in Appendix \ref{appendixB},
we can compute the required amplitudes. After standard operations we find 
in the closed string channel
\bea
{}_{\rm NS,U}\bra{D0} {\cal D} \ket{D0}_{\rm NS,U} &=&
{\alpha^\prime \pi \over 2}
{N_0^2 \over 2}V_{1} \Phi^4 (2\pi^2 \alpha^\prime)^{-5/2}\,
\int\limits_0^\infty d\ell \, \ell^{-5/2} \, \times
\nonumber \\
&& \hspace{-5cm}\times
\left\{
\left(\sum\limits_{n \in \Z} {\rm e}^{-\pi \ell {\alpha^\prime \over 2}
({n \over R})^2 } \right)^{4} 
+ \left(\sum\limits_{n \in \Z} {\rm e}^{-\pi \ell {\alpha^\prime \over 2}
({n \over R})^2 - 2i {yn \over R}} \right)^{4} 
\right\}
\, \left\{
\left({f_3 (q) \over f_1 (q)}\right)^8 - 
\left({f_4 (q) \over f_1 (q)}\right)^8 \right\} \, ,
\eea
with $q={\rm e}^{-\pi\ell}$. 
The closed string amplitudes related to the M${\ddot {\rm o}}$bius strips are
given by:
\bea
{}_{\rm NS,U}\bra{D0} {\cal D} \ket{C9}_{\rm NS,U} &=&
- {\alpha^\prime \pi \over 2}\, 
{T_0 \over 4\sqrt{2}} \, {T_9 \over 2} \, V_1 \,
2^5 2^{9/2} \, \int\limits_0^\infty d\ell \, \times \nonumber \\
&&\hspace{-4cm}\times \,
\left\{
\left( {f_1(iq) \over f_3(iq)}\right)
\left( {f_4(iq) \over f_2(iq)}\right)^{9}
-
\left( {f_1(iq) \over f_4(iq)}\right)
\left( {f_3(iq) \over f_2(iq)}\right)^{9} \right\} \, ,
\eea
and
\bea
{}_{\rm NS,U}\bra{D0} {\cal D} \ket{C5}_{\rm NS,U} &=&
{\alpha^\prime \pi \over 4}\,  {T_0 \over 4\sqrt{2}} \, {T_5 \over 2}
\, (2\pi R)^{-4} \, V_1 \, 2^{5/2} \, 
 \int\limits_0^\infty d\ell \, \times
\\
&&\hspace{-4cm}\times \,
\left\{
\left(\sum\limits_{n \in \Z} {\rm e}^{-\pi \ell {\alpha^\prime \over 2}
({n \over R})^2 - i{y n \over R}} \right)^{4} +
\left(\sum\limits_{n \in \Z} {\rm e}^{-\pi \ell {\alpha^\prime \over 2}
({n \over R})^2 + i{y n \over R}} \right)^{4} 
\right\} \, \times \nonumber
\\ 
&&\hspace{-4cm}\times \, \left\{
\left( {f_3(iq) \over f_1(iq)}\right)^{3}
\left( {f_4(iq) \over f_2(iq)}\right)^{5}
- \, 
\left( {f_4(iq) \over f_1(iq)}\right)^{3}
\left( {f_3(iq) \over f_2(iq)}\right)^{5} \right\} \, ,\nonumber
\eea
In order to obtain the amplitudes in open string channel
we perform a modular transformation,
$\ell= t^{-1}$,  for the cylinder and $\ell = (4t)^{-1}$ for the
M${\ddot {\rm o}}$bius strip. After using the modular properties of the 
functions\footnote{The explicit form
of the modular properties of these functions 
with imaginary argument can be found in \cite{Frau-nonbps}.}
$f_i$, we find:
\bea
{\cal A} &=&
V_{1} (8\pi^2 \alpha^\prime)^{-1/2} \, 
\int\limits_0^\infty {dt \over 2t} \, t^{-1/2} \, \times
\\
&& \times
\, \left\{
\left(\sum\limits_{m \in \Z} {\rm e}^{
- {2\pi t\over \alpha^\prime} (m R)^2 } \right)^{4} 
+
\left(\sum\limits_{m \in \Z} {\rm e}^{
- {2\pi t \over \alpha^\prime}R^2(m+ {y\over \pi R})^2 } \right)^{4} 
\right\}
\, \left\{
\left({f_3 (\tilde{q}) \over f_1(\tilde{q})}\right)^8 - 
\left({f_2 (\tilde{q}) \over f_1 (\tilde{q})}\right)^8 \right\} \, ,
\nonumber
\eea
with $\tilde{q}={\rm e}^{-\pi t}$.
\bea
{\cal M}_9 &=&
V_1 \, 2^4 \, (8\pi^2 \alpha^\prime)^{- 1/2} \,
\int\limits_0^\infty {dt \over 2t} \, t^{-1/2} \, \times 
\\
&& \times 
\, \left\{
{\rm e}^{- i {\pi \over 4}}
\left( {f_1(i\tilde{q}) \over f_4(i\tilde{q})}\right)
\left( {f_3(i\tilde{q}) \over f_2(i\tilde{q})}\right)^{9}
\, -  \, {\rm e}^{i {\pi \over 4}}
\left( {f_1(i\tilde{q}) \over f_3(i\tilde{q})}\right)
\left( {f_4(i\tilde{q}) \over f_2(i\tilde{q})}\right)^{9} \right\} \, ,
\nonumber
\eea
and
\bea
{\cal M}_5 &=&
V_1 \, 2^2 \, (8\pi^2\alpha^\prime)^{-1/2}\,
\int\limits_0^\infty {dt \over 2t} \, 
t^{-1/2} \, 
\left(\sum\limits_{m \in \Z} 
{\rm e}^{- { 8\pi t  \over \alpha^\prime} R^2(m + {y \over 2\pi R})^2} 
\right)^{4} \, \times
\\
&&\times \, \left\{
{\rm e}^{- i {\pi \over 4}}
\left( {f_4(i\tilde{q}) \over f_1(i\tilde{q})}\right)^{3}
\left( {f_3(i\tilde{q}) \over f_2(i\tilde{q})}\right)^{5}
\, - \,
{\rm e}^{i {\pi \over 4}}
\left( {f_3(i\tilde{q}) \over f_1(i\tilde{q})}\right)^{3}
\left( {f_4(i\tilde{q}) \over f_2(i\tilde{q})}\right)^{5} \right\} \, .
\nonumber
\eea
The open string spectrum for these non-BPS
branes in the type I orbifold is given by
the total amplitude ${\cal A}_{total} = {\cal A} + {\cal M}_9
+  {\cal M}_5 $.
These amplitudes can be checked to correspond to
the following amplitudes in the open string channel:
\bea
{\cal A} &=& \int\limits_0^\infty {dt \over 2t} 
{\rm Tr}_{NS,R}\left({1 \over 4}(-1)^{F_s} {\rm e}^{-2\pi t
L_0}\right) \, , 
\nonumber \\
{\cal M}_{9} &=&
\int\limits_0^\infty {dt \over 2t} 
{\rm Tr}_{NS}\left( {1 \over 4} \O \, {\rm e}^{-2\pi t L_0}\right) 
\, ,\nonumber \\
{\cal M}_{5} &=&
\int\limits_0^\infty {dt \over 2t} 
{\rm Tr}_{NS}\left({1 \over 4}
{\cal I}_4 \O \, {\rm e}^{-2\pi t L_0}\right)  \, ,
\eea
where besides the trace over the oscillators, the Tr in each case includes
the trace of Chan-Paton factors,
integration over momenta in non-compact directions and summing over
discrete momenta and winding states. The total amplitude can therefore
be written in the following compact form:
\be
{\cal A}_{total} = \int\limits_0^\infty {dt \over 2t} 
{\rm Tr}_{NS,R}
\, \left\{ (-1)^{F_s} 
\, \left( {1 + {\cal I}_4 \over 2} \right)
\left( {1 + \O \over 2} \right)
\, {\rm e}^{-2\pi t L_0} \right\} \, . 
\ee
In this expression one must take into account that ${\cal I}_4$
in the NS sector yields zero and in the R sector ${\cal I}_4$,
$\O$ and ${\cal I}_4 \O$ are zero. Note that it is crucial to take the linear
combination of the brane and its image as the boundary state to get the nontrivial
vanishing result for the ${\cal I}_4$ projection while still getting a non-vanishing
contribution for ${\cal I}_4\O$. This is essential for the closed-open
consistency condition for the orbifold theory. This also reflects that we need to
include the C-5 state to account for the interaction between the brane and the O5
planes.

We can now check that the tachyon state cancels out from the open
string spectrum by taking the limit $t \to \infty$.
In this limit we obtain the following leading terms in the amplitude:
\bea
{\cal A} &\simeq&
V_1 (8\pi^2 \alpha^\prime)^{-1/2}
\int {dt \over 2t} t^{- 1/2} \, \tilde{q}^{-1} \, \left(
1 + {\rm e}^{- {8t \over \pi \alpha^\prime }{y^2} } \right) \, ,
\nonumber \\
{\cal M}_9 &\simeq&
- V_1 (8\pi^2 \alpha^\prime)^{-1/2}
\int {dt \over 2t} t^{-1/2} \, \tilde{q}^{-1} \, ,
\\
{\cal M}_5 &\simeq&
- V_1 (8\pi^2 \alpha^\prime)^{-1/2}
\int {dt \over 2t} t^{-1/2} \, \tilde{q}^{-1} \, \left(
{\rm e}^{- {8t \over \pi \alpha^\prime }{y^2} } \right) \, ,\nonumber
\eea
hence ${\cal A}_{total} \simeq 0$, and the tachyon cancels.
It is important to note that the contribution from the D-particle$-$C5 amplitude
cancels with
the winding (y-dependent) part in the particle$-$image-particle amplitude and the
contribution from the particle$-$C9 amplitude cancels with the y-independent part.
The reason for such a mechanism is that the O9 plane, being space-time
filling, is not sensitive to the position of the brane. On the contrary, 
the C5-state is localised at the origin, hence it is sensitive to the
position of the brane. This mechanism will be relevant in the next subsection to
find other stable non-BPS $\Z_2$-charged branes.

Note that there are also open strings that stretch between the
D-particle and the background D5 and D9 branes. However, these do not have
any tachyon because the intercept in the NS-sector for these strings 
is strictly negative since 
\be
a_{NS} = {1 \over 2} - {{ND} \over 2} \, ,
\ee
where $ND$ is the number of mixed Dirichlet - Neumann directions and is 9 and 5 for
D0-D9 and D0-D5 strings respectively. Thus the D-particle in this orbifold theory is
absolutely stable as it is in the case before orbifolding.

%%%%%%%%%%%%%%%%%%%%%%%%%%%%%%%%%%%%%%%%%%%%%%%%%%%%%%%%%%%%%%%%%%

\subsection{Stable Non-BPS Branes with $\Z_2$ Charge}

We now generalise the previous results for a generic $\Z_2$ brane. 
We note that $(r,s)$ branes with $s$ different from zero 
cannot be properly defined as $\Z_2$-charged branes. The reason is
that for non-zero $s$ we are able to distinguish a brane and its image only
by introducing a Wilson line. This is consistent with T-duality but
requires, however, the brane to have a twisted 
sector\footnote{The Wilson line creates a flux on the world
volume. In the presence of the orbifold, since the
string closes only up to a phase, there is a flux deficit. 
In order to compensate for this a twisted sector must be then
introduced.}, i.e. the brane
must be either a fractional or ${\Z}$-charged.
Thus only $(r,0)$ branes can be formulated as $\Z_2$ charged branes.
The open-closed
duality consistency works for these branes just as the case for the D-particle and
we will not repeated again. 
The stability analysis follows closely to the procedure used
in \cite{Frau-nonbps} for the uncompactified theory. We introduce a 
parameter $m_r$ to renormalise the tension
of these branes, which will be fixed by imposing that 
there are no tachyons in the open string spectrum on that brane which is also
consistent with open-closed consistency as we will see below.
In order to achieve this, 
we calculate the interaction between two of these non-BPS branes and
translate the result to open string channel. Then, we impose 
that there are no tachyons in the open string spectrum, which will
fix the parameter $m_r$. Only positive values of $m_r$ will be considered
to produce a consistent stable brane. 
  
In the orientifold, the $\Z_2$-charged branes are thus represented as follows:
\be
{\ket{D(r,0)}}_{T^4/\Z_2} =
{1 \over \sqrt{2}} \left( \ket{D(r,0)}_{\rm NS,U} 
+ \ket{C9} + \ket{C5} \right) \, .
\ee
where, as before, the state $\ket{D(r,0)}_{\rm NS,U}$ represents the
the NS-NS sector of the brane in type IIB on $T^4/{\cal I}_4$, hence
\bea
\ket{D(r,0)}_{\rm NS,U} &=& 
{1 \over 2\sqrt{2}}\left( \ket{D(r,0),a^\alpha,y^i,+}_{\rm NS,U} +
\ket{D(r,0),a^\alpha,-y^i,+}_{\rm NS,U} \right. \\ 
&&\hspace{2cm} \left.- \ket{D(r,0),a^\alpha,y^i,-}_{\rm NS,U}  -
\ket{D(r,0),a^\alpha,-y^i,-}_{\rm NS,U}  \right) \, ,\nonumber
\eea
with $a^\alpha$ being the position of the brane in the $(5-r)$ non-compact
transverse directions and $y^i$ is that of the $4$
compact transverse directions.  
The relevant open string amplitudes are given by
the annulus and M${\ddot {\rm o}}$bius amplitudes, which can be
obtained from the closed channel as before:
\bea
{\cal A} &=& {1 \over 2} {}_{\rm NS,U}\bra{D(r,0)} {\cal D} 
\ket{D(r,0)}_{\rm NS,U} \, ,\\
{\cal M}_9 + {\cal M}_5 &=& {1 \over 2} {}_{\rm NS,U}\bra{D(r,0)} {\cal D} 
\ket{C9} + 
{1 \over 2} {}_{\rm NS}\bra{D(r,0)} {\cal D} 
\ket{C5} + c.c. \, .\nonumber
\eea
We normalise the brane in the following way:
\be
\ket{D(r,0),\eta}_{\rm NS,U} = N_p \,
\ket{D(r,0)_X}\ket{D(r,0)_\psi,\eta}_{\rm NS,U}\ket{D(r,0)_{gh}}
\ket{D(r,0)_{sgh}}_{\rm NS} \, ,
\ee
with
\be
N_r = m_r {T_r \over 2} \left({2\pi R \over \Phi} \right)^2
(2\pi R)^{-4} \, ,
\ee
where $T_r$ is the tension of a BPS D$r$-brane in type II theory:
\be
T_r = \sqrt{\pi} (2\pi \sqrt{\alpha^\prime})^{3-r} \, .
\ee
Moreover, the bosonic
sector is modified as follows:
\bea
\ket{D(r,0)_X} &=& \delta^{(5-r)}(q^\alpha - a^\alpha)
\, 
%\left(\sum\limits_{m \in \Z} {\rm e}^{iq_m { m R \over
%\alpha^\prime}}
%\right)^s
%\,
\left(\sum\limits_{n \in \Z} {\rm e}^{i q_n {n \over R}}
\right)^{4}
\, \times
\\
&&\,\times \, \prod\limits_{n=1}^{\infty}
{\rm e}^{-{1 \over n} \alpha_{-n} \cdot {\cal S} \cdot
\tilde{\alpha}_{-n}}
\, \ket{k,m,n=0} \, . \nonumber
\eea
With this information and using the 
explicit form of the crosscaps given in Appendix \ref{appendixB},
we can compute the required amplitudes. 
We obtain the corresponding amplitudes in the open channel
with the usual modular transformations. After these operations we find
in the open channel:
\bea
{\cal A} &=&
{1\over 2} (m_r)^2 \, V_{r+1} \, (8\pi^2 \alpha^\prime)^{-({r+1 \over 2})} 
%\Phi^4 (2\pi^2 \alpha^\prime)^{-({5-r \over 2})} \,
%R^{4-2s} (\alpha^\prime)^{-2+s} \, 
\int\limits_0^\infty {dt \over 2t} \, t^{-({r+1 \over 2})} 
\, \left\{
\left({f_3 (\tilde{q}) \over f_1(\tilde{q})}\right)^8 - 
\left({f_2 (\tilde{q}) \over f_1 (\tilde{q})}\right)^8 \right\} 
\, %\times
%\nonumber
\\
&\times&
%\left(\sum\limits_{n \in \Z} {\rm e}^{-2\pi t {\alpha^\prime}
%({n \over R})^2 } \right)^{s}
\left\{
\left( \sum\limits_{m \in \Z} {\rm e}^{-2\pi t
{(m R)^2 \over \alpha^\prime}} \right)^{4} \,
 + \, \left( \sum\limits_{m \in \Z} {\rm e}^{-2\pi t
{R^2 \over \alpha^\prime} (m + {y \over \pi R})^2} \right)^{4} \,
\right\} \, ,\nonumber
\eea
with $\tilde{q}={\rm e}^{-\pi t}$.
\bea
{\cal M}_9 &=&
- m_r \, V_{r+1} \, 2^{{7-r \over 2}} \, (8\pi^2 \alpha^\prime)^{-({r+1 \over 2})}
\, \int\limits_0^\infty {dt \over 2t} \, 
t^{-({r+1 \over 2})} \, 
%\left(\sum\limits_{n \in \Z} {\rm e}^{-8\pi t \alpha^\prime  
%({n \over R})^2} \right)^{s} \, 
%\times
\\
&\times &\, \left\{
{\rm e}^{i {\pi \over 4} (r-5)}
\left( {f_4(i\tilde{q}) \over f_1(i\tilde{q})}\right)^{r-1}
\left( {f_3(i\tilde{q}) \over f_2(i\tilde{q})}\right)^{9-r}
-\, %\left.
{\rm e}^{- i {\pi \over 4} (r-5)}
\left( {f_3(i\tilde{q}) \over f_1(i\tilde{q})}\right)^{r-1}
\left( {f_4(i\tilde{q}) \over f_2(i\tilde{q})}\right)^{9-r} \right\} \, ,
\nonumber
\eea

and

\bea
{\cal M}_5 &=&
m_r \, V_{r+1} \, 2^{{3 - r - s\over 2 }} \, (8\pi^2 \alpha^\prime)^{-({r+1
\over 2})} \, \int\limits_0^\infty {dt \over 2t} \, 
t^{-({r+1 \over 2})} \, 
\left(\sum\limits_{m \in \Z} 
{\rm e}^{- 8\pi t 
{R^2 \over \alpha^\prime} (m + {y \over 2\pi R})^2} \right)^{4} \, 
%\times
\\
\nonumber 
&\times &\, \left\{
{\rm e}^{- i {\pi \over 4} (1-r)}
\left( {f_4(i\tilde{q}) \over f_1(i\tilde{q})}\right)^{3+r}
\left( {f_3(i\tilde{q}) \over f_2(i\tilde{q})}\right)^{5-r}
\, \right. \\
&&\hspace{3cm} \, - \, \left.
{\rm e}^{i {\pi \over 4} (1-r)}
\left( {f_3(i\tilde{q}) \over f_1(i\tilde{q})}\right)^{3+r}
\left( {f_4(i\tilde{q}) \over f_2(i\tilde{q})}\right)^{5-r} \right\} \, .
\nonumber
\eea
As was the case for the D-particle, in the previous section, to test
the cancellation of the tachyons we
make an expansion in  powers of $\tilde{q}$, for $t \to \infty$ and 
we obtain the following leading term in the  amplitude 
${\cal A}_{total} = {\cal A} + {\cal M}_9 + {\cal M}_5$:
%+ {\cal M}_9^* + {\cal M}_5 + {\cal M}_5^*$:
\bea
{\cal A}_{total} &\simeq&
{m_r\over 2} \,V_{r+1} \, (8 \pi^2 \alpha^\prime)^{- ({r+1 \over 2})} \, \int 
{dt \over 2t} t^{- ({r+1 \over 2})} \, 
\tilde{q}^{-1} \, %\times 
\label{extract}
\\
&\times &\, \left\{\left(
m_r \, - \, 2 \, {\rm sin}\, [{\pi \over 4}(r-5)]\right)
\, + \,\left( m_r \, - \, 2 \, {\rm sin}\, [{\pi \over 4}(1-r)]\right) 
{\rm e}^{- 8 t {y^2 \over \alpha^\prime \pi}}  \, 
\right\}\, ,
\nonumber
\eea
which corresponds to the contribution from the tachyon.
In order that the tachyons are projected out from the spectrum 
this contribution must vanish, hence the following condition must be satisfied:
\be
m_r = 2 \, {\rm sin}\, [{\pi \over 4}(1-r)]
\, .
\label{no-tachyon-condition}
\ee
Note that if we take $r = 0$, which corresponds to the D-particle,
we get $m_0 = {\sqrt 2}$, which is the normalisation factor we have found 
for the tension of the D-particle using open-closed consistency condition. 
Besides the D-particle, we find that the only other stable non-BPS
brane is the $(-1,0)$-brane, i.e. a D-instanton for which 
we have $m_{-1} = 2$. For all other
possible values of $r$ ($=1,2,4,5$) $m_r$ is either zero or negative, hence 
they are not allowed. It is worth noting that this D-instanton, 
having the mass twice that of
the BPS D-instanton of the Type II theory but with zero RR charge, 
is possibly just 
the superposition of the instanton and the anti-instanton. In the 
next section we will find another instanton but carrying a 
twisted RR charge. It may well be two
different descriptions of the same instanton just like the familiar stable
D-particle in Type IIB in the orbifold $T^4/(I_4 (-1)^{F_L})$.

We now analyse the tachyons in the strings stretching to the
tadpole-cancelling branes. For convenience, we consider the case for 
general $(r,s)$ branes, keeping in mind that for $\Z_2$ branes only
the case $s=0$ is applicable.
For the strings stretching between the
non-BPS brane and the D9 brane we have
\be
M^2 = \sum\limits_{s \, (NN) \, directions} \left( {n_i \over R_i} \right)^2
+ {1 \over \alpha^\prime} \left( {5-p \over 8} \right) \, .
\label{p-9-tachyons}
\ee
For $p = r+s \leq 5$ there are no tachyons, for any radius. 
Since the ground state of this sector ($n_i = 0$, $\forall i$)
is not projected out of the spectrum, 
when $p > 5$ the brane contains always at least one tachyon.
%there is a critical radius below which the brane is stable:
%\be
%R \leq R_c = 2 \sqrt{ 2\alpha^\prime \over p - 5 } \, .
%\ee
%Above this critical radius the brane is unstable.
On the other hand, for the strings stretching between the
non-BPS brane and the D5 brane we have
\be
M^2 = \sum\limits_{4-s \, (DD) \, directions} 
\left( {m_i R_i \over \alpha^\prime} \right)^2
+ {1 \over \alpha^\prime} \left( {1-r+s \over 8} \right) \, .
\label{p-5-tachyons}
\ee
For $r-s \leq 1$ there are no tachyons, for any radius. 
Similarly, for $r-s > 1$ the brane contains at least one 
tachyon in this sector.
%there is a critical radius above which the brane is stable:
%\be
%R \geq R_c = {1 \over 2} \sqrt{ {\alpha^\prime \over 2} (r-s-1)} \, .
%\ee
%Below this critical radius the brane is unstable.
For $\Z_2$ branes this implies that there are no tachyons in any of
these sectors if $r \leq 1$, hence
the D-particle $(0,0)$ and the D-instanton $(-1,0)$ are
absolutely stable non-BPS branes with $\Z_2$ charge.

%%%%%%%%%%%%%%%%%%%%%%%%%%%%%%%%%%%%%%%%%%%%%%%%%%%%%%%%%%%%%%%%%%%%%%

\subsection{Stable Non-BPS Branes with $\Z$ Charge}

%%%%%%%%%%%%%%%%%%%%%%%%%%%%%%%%%%%%%%%%%%%%%%%%%%%%%%%%%%%%%%%%%%%%%%

These are the branes that carry a twisted R-R sector, namely
\be
r=-1,3 \, ,\qquad s=0,1,3,4 \, ,
\ee
and they have the following form:
\be
{\ket{D(r,s)}}_{T^4/\Z_2}
= {1 \over \sqrt{2}} \left( \ket{D(r,s)}_{\rm NS,U} + \ket{D(r,s)}_{\rm R,T}
+ \ket{C5} + \ket{C9} \right) \, .
\ee 
Note that we do not take the combination of brane
and its image, rather we use the formal definition and we localize the branes
at the origin. As for the truncated branes in a type IIB orbifold, they
may be stable for some range of the radii.
As before, we can obtain the amplitude of open strings on one 
of these branes using the boundary states:
\bea
{\cal A}_{total} &=& 
{}_{T^4/\Z_2}{\bra{D(r,s)}} {\cal D} {\ket{D(r,s)}}_{T^4/\Z_2}
\nonumber \\
&=& {\cal A}_{NS-NS} + {\cal A}_{R-R,T} + {\cal M}_{9, NS-NS} +
{\cal M}_{5, NS-NS} 
\eea
The correspondence between the spin structures of the loop and tree 
channels is as follows:
\bea
{\cal A}_{NS-NS} =
\int\limits_0^\infty {dt \over 2t} 
{\rm Tr}_{NS,R}\left({1 \over 4}(-1)^{F_s} {\rm e}^{-2\pi t L_0}\right)  
&=& 
{1 \over 2}{}_{\rm NS,U}{\bra{D(r,s)}} {\cal D} {\ket{D(r,s)}}_{\rm NS,U} \, ,
\nonumber \\
{\cal A}_{R-R,T} = 
\int\limits_0^\infty {dt \over 2t} 
{\rm Tr}_{NS}\left( {1 \over 4}{(-1)^{F+G} {\cal I}_4} 
\, {\rm e}^{-2\pi t L_0}\right)  
&=& 
{1 \over 2}{}_{\rm R,T}{\bra{D(r,s)}} {\cal D} {\ket{D(r,s)}}_{\rm R,T} \, ,
\nonumber \\
{\cal M}_{9, NS-NS} =
\int\limits_0^\infty {dt \over 2t} 
{\rm Tr}_{NS}\left( {1 \over 4} \O \, {\rm e}^{-2\pi t L_0}\right)  
&=& 
{1 \over 2}{}_{\rm NS,U}{\bra{D(r,s)}} {\cal D} {\ket{C9}}_{\rm NS,U}  
\, + \, c.c. \, ,
\nonumber \\
{\cal M}_{5, NS-NS} =
\int\limits_0^\infty {dt \over 2t} 
{\rm Tr}_{NS}\left({1 \over 4}
(-1)^{F+G} {\cal I}_4 \O \, {\rm e}^{-2\pi t L_0}\right)  
&=& 
{1\over 2}{}_{\rm NS,U}{\bra{D(r,s)}} {\cal D} {\ket{C5}}_{\rm NS,U} 
\, + \, c.c. \, ,
\nonumber
\eea
where $F_s$ is the space-time fermion number and $(-1)^{F+G}$ is the GSO
projection operator including the super-ghosts.
Furthermore, we take into account that
\be
{\rm Tr}_{R}\left( (-1)^{F+G} {\cal I}_4 \, {\rm e}^{-2\pi t
L_0}\right)  =0 \, ,
\ee
due to the presence of zero modes in the R-vacuum, and
\be
 {\rm Tr}_{R}\left(\O \, {\rm e}^{-2\pi t L_0}\right) =0 \, ,
\ee
since $\O$ acts as a product of gamma-matrices  on the R-vacuum,
and finally 
\be
{\rm Tr}_{R}\left((-1)^{F+G} {\cal I}_4 \O \, {\rm e}^{-2\pi t L_0}\right)  
= 0 \, ,
\ee
since although
the action of $\O$ and ${\cal I}_4$ compensate
it still vanishes due to $(-1)^{F+G}$.
Accordingly we can write the total amplitude in terms of an open string
loop as follows:
\be
{\cal A}_{total} =
\int\limits_0^\infty {dt \over 2t}
{\rm Tr} \left\{ (-1)^{F_s} \, \left( {1 + (-1)^{F+G} {\cal I}_4 \over
2}\right)
\left( 1 + \O \over 2 \right)
 {\rm e}^{-2\pi t L_0} \right\} \, .
\ee
The open strings on these D-branes have winding and Kaluza-Klein
modes
\be
  M^2 = 
\sum_{4-s (DD) directions} \left({m R \over
\alpha^\prime}\right)^2 +
\sum_{s (NN) directions} \left({n \over
R }\right)^2 +
{1 \over \alpha^\prime} \left( N- {1 \over 2} \right) \, ,
\ee
hence for a certain range of the radii there are no tachyons
in the spectrum: $R \geq \sqrt{\alpha^\prime /2}$ for the
 DD-directions and $R \leq \sqrt{2\alpha^\prime}$ for the
NN-directions.
Moreover, at particular radii, the tachyon modes become massless.
In the case of non-BPS branes in type II orbifolds, these critical
radii coincide with the critical radii where the brane has a
vanishing 1-loop amplitude \cite{Gaberdiel-Sen}. 
In this case, however, this is not possible since 
there is a remaining NS-NS interaction of the brane with the crosscaps
that cannot be cancelled.

In order to carry out the analysis of
tachyons in the strings stretched from the non-BPS brane to the
tadpole-cancelling branes we can use the
analysis carried out before for the $\Z_2$-charged non-BPS branes.
Using (\ref{p-9-tachyons}) and (\ref{p-5-tachyons})
one can see that the {\it instantonic} branes 
${\ket{D(-1,0)}}_{T^4/Z_2}$,
${\ket{D(-1,1)}}_{T^4/Z_2}$,
${\ket{D(-1,3)}}_{T^4/Z_2}$
and ${\ket{D(-1,4)}}_{T^4/Z_2}$
have no tachyons in this sector at any radius. 

On the other hand, the branes 
${\ket{D(3,0)}}_{T^4/Z_2}$, ${\ket{D(3,1)}}_{T^4/Z_2}$,
${\ket{D(3,3)}}_{T^4/Z_2}$ and ${\ket{D(3,4)}}_{T^4/Z_2}$ contain
at least a tachyon in this sector.
Accordingly, the {\it instantonic} truncated branes listed above
are fully stable if the radii take the values 
$R \geq \sqrt{\alpha^\prime /2}$ for the
DD-directions and $R \leq \sqrt{2\alpha^\prime}$ for the
NN-directions, for each particular brane.
Finally, we observe that the D-instanton of type I,
which becomes a truncated brane with twisted R-R charge in the type I orbifold,
can be stable in such theory.

%On the other hand, 
%${\ket{D(3,0)}}_{T^4/Z_2}$ has no tachyons
%for $R_i>\sqrt{\alpha^\prime}/2$, $i=6,7,8,9$;
%${\ket{D(3,1)}}_{T^4/Z_2}$ has no tachyons
%for $R_i \geq \sqrt{\alpha^\prime}/2\sqrt{2}$, $i=7,8,9$, for any
%$R_6$;
%${\ket{D(3,3)}}_{T^4/\Z_2}$ has no tachyons
%if $R_i \leq 2\sqrt{2\alpha^\prime}$, $i=6,7,8$, for any $R_9$;
%and 
%${\ket{D(3,4)}}_{T^4/\Z_2}$ has no tachyons in this sector
%if $R_i \leq 2\sqrt{\alpha^\prime}$, $i=6,7,8,9$.
%Notice that these ranges lie within the values for which there are
%tachyons in the DD- and NN- directions, as given above. Thus they
%do not imply any further restrictions to the radii for stability.
%Accordingly, we can have any of the truncated branes listed above
%to be stable if the radii take the values 
%$R \geq \sqrt{\alpha^\prime /2}$ for the
% DD-directions and $R \leq \sqrt{2\alpha^\prime}$ for the
%NN-directions, for that particular brane.

%%%%%%%%%%%%%%%%%%%%%%%%%%%%%%%%%%%%%%%%%%%%%%%%%%%%%%%%%%%%%%%%%%%%%%%%%

\section{Conclusions}
\label{section5}

In this article we have presented a thorough analysis of the
action of the $\Omega$ projection on the boundary states of type IIB
theory in a $T^4/{\cal I}_4$ orbifold. Of particular interest are the
R-R sectors and the twisted NS-NS sector.
We have shown how to derive the action of $\Omega$ in the
R-R sectors of the covariant boundary states, 
which are formulated in the
asymmetric picture $(-1/2,-3/2)$, 
for which the superghost zero-modes
play an important role. In the twisted NS-NS sector, where
there are no superghost zero modes, we
have implemented the action of $\Omega$ through the algebra of
fermionic zero modes. Since these results rely on basic features of
the covariant boundary states, which are common to other theories, 
our approach could be extended
to other orbifold and orientifold theories in order to derive
their complete D-brane spectrum in a systematic way.

Using the results of our analysis of the $\Omega$ projection, 
and taking into account the
action of the orbifold on the different sectors, we have derived which
boundary states are present in the type I orbifold. From this we 
have derived the spectrum of BPS and non-BPS D-branes 
of type I on $T^4/{\cal I}_4$.
Regarding the non-BPS D-branes, they are divided 
into $\Z$-charged branes, which have 
a twisted R-R sector, and $\Z_2$-charged branes, which are described
by a untwisted NS-NS sector only.

The analysis of the stability of these non-BPS D-branes
has been carried out. For the $\Z_2$-charged non-BPS 
branes we have found that only branes of the type $(r,0)$ can be
formulated as such.  
%that the non-BPS D-particle of type I remains
%stable in the orbifold. We also have found other branes 
%which have no tachyons in their spectrum, namely:
%$\ket{D(0,0)}$, $\ket{D(5,1)}$, $\ket{D(4,2)}$,
%$\ket{D(2,4)}$, $\ket{D(5,3)}$, $\ket{D(4,4)}$, $\ket{D(5,2)}$, 
%$\ket{D(4,3)}$, $\ket{D(-1,0)}$.
%However, we have found 
From those, only the particle
$\ket{D(0,0)}$ and the instanton $\ket{D(-1,0)}$ are
absolutely stable since they do not contain any tachyon
in their spectrum of open strings.

On the other hand, the non-BPS $\Z$-charged branes 
have no tachyons on their brane spectrum for a particular range of the
radii, namely, $R \geq \sqrt{\alpha^\prime /2}$ for the
DD-directions and $R \leq \sqrt{2\alpha^\prime}$ for the
NN-directions, as occurs in the type IIB orbifold. 
We have found that, moreover, only the branes
${\ket{D(-1,0)}}$, ${\ket{D(-1,1)}}$, ${\ket{D(-1,3)}}$, ${\ket{D(-1,4)}}$, 
do not have any tachyons in the 
open strings stretching to the tadpole cancelling branes, 
so they can be considered as fully stable for the
range of the radii mentioned above.
%The other branes, namely, ${\ket{D(3,0)}}$, 
%${\ket{D(3,1)}}$, ${\ket{D(3,3)}}$, ${\ket{D(3,4)}}$,
%have tachyons in the open strings stretching to the tadpole cancelling branes.

It is interesting to analyse the fate of the well-known non-BPS branes
of type I theory after the orbifolding.
The non-BPS D-particle becomes a stable non-BPS $\Z_2$-charged brane.
D7 and D8 branes cannot be formulated as $\Z_2$-charged branes, and
the D8 brane cannot carry any R-R charge, hence there are no D8-branes
in the type I orbifold.
There is, however, a non-BPS D$7$ branes with twisted R-R charge for the
orientation $(3,4)$, but it is unstable since it has tachyons
in the open strings stretching to the tadpole-cancelling branes.
Finally, the non-BPS D-instanton of type I becomes a stable non-BPS
brane and appears as a $\Z$-charged brane when it sits at any of the 
fixed planes and all radii fulfill $R \geq \sqrt{\alpha^\prime /2}$, 
and as a $\Z_2$-charged brane when it is separated from the 
fixed planes.

It would be of great interest to extend the techniques used in this article
to other orientifolds where the non-BPS branes play a more
relevant role in the consistency of the theory
\cite{Bianchi-1,Bianchi-2,more-nonbps} and in particular
to those models with a phenomenological interest \cite{Aldazabal-etal}.

%%%%%%%%%%%%%%%%%%%%%%%%%%%%%%%%%%%%%%%%%%%%%%%%%%%%%%%%%%%%%%%%%%%%%%%%%

\section*{Acknowledgements}

We are grateful to Ashoke Sen and Matthias Gaberdiel for many discussions
and for their useful comments on a draft of the paper.
S.P. also thanks S. Mukhi and K.S. Narain for many useful discussions.
The work of E.E. is supported by the European Community program
{\it Human Potential} under the contract HPMF-CT-1999-00018.
This work is also partially supported by the PPARC grant
PPA/G/S/1998/00613.

%%%%%%%%%%%%%%%%%%%%%%%%%%%%%%%%%%%%%%%%%%%%%%%%%%%%%%%%%%%%%%%%%%%%%%%%%
%%%%%%%%%%%%%%%%%%%%%%%%%%%%%%%%%%%%%%%%%%%%%%%%%%%%%%%%%%%%%%%%%%%%%%%%%
%%%%%%%%%%%%%%%%%%%%%%%%%%%%%%%%%%%%%%%%%%%%%%%%%%%%%%%%%%%%%%%%%%%%%%%%%

  \appendix
  \section{$\O$ and the Asymmetric Picture}
  \label{appendixA}

In this appendix we give a detailed computation of the
action of $\Omega$ on the boundary states formulated in the
asymmetric superghost picture $(-1/2,-3/2)$.
The action of $\Omega$ on the (untwisted) R-R boundary states in the 
light-cone gauge formulation was derived before
in \cite{Bergman-Gaberdiel-open}. However, only the covariant formulation
\cite{covariant-state} allows the description of D-branes which are of type 
spacetime-filling or domain-wall with 
respect to either the bulk or the orbifold fixed planes,
hence the relevance of understanding the action of $\Omega$ in this
formulation. For instance, in the present
case, it allows the description of the untwisted R-R sector of a D9-brane 
and the twisted R-R sector of any D-brane for which all the spatial
directions of the orbifold fixed plane are Neumann-directions. 

Usually, in order to find out which (BPS) D-branes
survive the $\O$ projection, one 
may consider the R-R state in the $(-1/2,-1/2)$ picture:
\be
\ket{F_{p+2}} = {1 \over (p+2)!} \,
F_{\mu_0 \dots \mu_{p+1}} \left({\cal C}_{(10)} \Gamma^{\mu_0 \dots \mu_{p+1}} 
\right)_{AB} \, \ket{A}_{-1/2} \otimes \widetilde{\ket{B}}_{-1/2}
\, .
\ee
Applying $\Omega$ and taking into account that for the R-R vacuum
\be
\O \left( \ket{A}_{-1/2} \otimes \widetilde{\ket{B}}_{-1/2} 
\right) = \widetilde{\ket{A}}_{-1/2} \otimes {\ket{B}}_{-1/2}
=  - \ket{B}_{-1/2} \otimes \widetilde{\ket{A}}_{-1/2} \, ,
\ee
we find:
\be
\O \, \ket{F_{p+2}} = - (-1)^{{1 \over 2}(p+1)(p+2)} \,  \ket{F_{p+2}} \, ,
\ee
where we have used that $p=odd$ for type IIB and that
${\cal C}_{(10)}^T=-{\cal C}_{(10)}$.
This state is left invariant for $p=1,5$, and changes sign for
$p=-1,3,7$.
Notice that here it is crucial that the picture is left-right symmetric,
in order to recover the original ordering of the pictures.

On the other hand,  the boundary state of the R-R sector
in the covariant formulation is constructed upon the
vacuum in the asymmetric picture $(-1/2,-3/2)$ \cite{Divecchia-etal}. 
This boundary state is given by\footnote{For simplicity, we omit
the label $(r,s)$ in what follows.}
\be
\ket{D}_{\rm R,U} =
{1 \over 2}\left( \ket{D,+}_{\rm R,U} +
\ket{D,-}_{\rm R,U} \right) \, ,
\ee
with
\be
\ket{D,\eta}_{\rm R,U} =
{Q \over 2} \, \ket{D_X}
\ket{D_\psi,\eta}_{\rm R,U} \ket{D_{gh}} 
\ket{D_{sgh},\eta}_{\rm R} \, ,
\ee
and
\bea
\ket{D_{sgh},\eta}^{(0)}_{\rm R} 
\ket{D_\psi,\eta}^{(0)}_{\rm R,U} &=&
{\rm e}^{i\eta \gamma_0 \tilde{\beta}_0} 
{\cal M}_{AB} \, \ket{A}_{-1/2} \otimes \widetilde{\ket{B}}_{-3/2} \, ,\\
{\cal M}_{AB}(\eta) &=& 
\left( {\cal C}_{(10)} 
\Gamma^0 \cdots \Gamma^p {1 + i\eta \Gamma_{11} \over
1+ i\eta}\right)_{AB} \, ,\qquad p=r+s \, .\nonumber 
\eea
However, under the action of $\Omega$, the superghost pictures
are interchanged:
\be
\O \left( \ket{A}_{-1/2} \otimes \widetilde{\ket{B}}_{-3/2} 
\right) = \widetilde{\ket{A}}_{-1/2} \otimes {\ket{B}}_{-3/2}
=  - \ket{B}_{-3/2} \otimes \widetilde{\ket{A}}_{-1/2} \, .
\label{asymmetric-twist}
\ee
In this case, in order to bring it back to the
$(-1/2,-3/2)$ picture, a more careful treatment is needed.
In order to understand how $\Omega$ acts in the states with asymmetric
picture 
we introduce the formal operators \cite{Callan-bis}
\be
\delta(\gamma_m) \, ,\qquad \delta(\beta_n) \, ,
\ee
for each of the holomorphic sectors,
which allow to relate different pictures
\be
\delta( \beta_{-q - 3/2}) \ket{q} = \ket{q+1} \, ,\qquad
\delta( \gamma_{q + 1/2}) \ket{q} = \ket{q-1} \, .
\ee
Moreover, they have the following commutation relations:
\bea
\lbrack \beta_m,\delta (\beta_n) \rbrack = 0 \, &,& 
\qquad \lbrack \gamma_m, \delta(\beta_n) \rbrack 
= \delta_{n, -m} {d \over d \beta_n} \delta(\beta_n) \, ,
\nonumber \\
\lbrack \gamma_m ,\delta (\gamma_n) \rbrack = 0 \, &,& 
\qquad \lbrack \beta_m, \delta(\gamma_n) \rbrack 
= - \delta_{n, -m} {d \over d\gamma_n} \delta(\gamma_n) \, .
\nonumber
\eea
Therefore, the two possible vacuum states of the superghost zero-modes
\be
\beta_0 \ket{\downarrow} = 0 \, ,\qquad \gamma_0 \ket{ \uparrow} =0 \, ,
\ee
can be characterised as follows:
\be
\delta(\beta_0) \ket{\uparrow} = \ket{\downarrow} \, ,\qquad
\delta(\gamma_0) \ket{\downarrow} = \ket{ \uparrow} \, ,
\ee
in analogy with the two vacuum states of the ghost zero-modes.
On the other hand, considering the
overlapping conditions for the zero-mode 
of the superghost boundary state
\be
(\gamma_0 + i\eta {\tilde \gamma}_0) \ket{D_{sgh}, \eta}^{(0)}_{\rm R}
= 0 \, ,\qquad
(\beta_0 + i\eta {\tilde \beta}_0) \ket{D_{sgh}, \eta}^{(0)}_{\rm R}
= 0 \, ,
\ee
we can find the following solution to these conditions
\be
\ket{D_{sgh}, \eta}^{(0)}_{\rm R} = 
\delta(\gamma_0 + i\eta {\tilde \gamma}_0) \, 
\ket{ \downarrow } \otimes \widetilde{ \ket{ \downarrow } }\, .
\ee
Using the above results, we can rewrite the zero-mode part of the superghost 
boundary state in the asymmetric $(-1/2,-3/2)$ picture
as follows:
\bea
\ket{D_{sgh}, \eta}^{(0)}_{\rm R} &=&
{\rm e}^{i\eta \gamma_0 {\tilde \beta}_0} \, \ket{-1/2} \otimes 
\widetilde{\ket{-3/2}}\nonumber \\
&=& 
{\rm e}^{i\eta \gamma_0 {\tilde \beta}_0} 
\, \delta({\tilde \gamma}_0) \, 
\ket{-1/2} \otimes \widetilde{\ket{-1/2}} \nonumber \\
&=& 
\delta({\tilde \gamma_0} -i\eta \, \gamma_0) 
\, \ket{-1/2} \otimes \widetilde{\ket{-1/2}} \, ,
\eea
which allows to easily derive its transformation rule
under $\Omega$:
\bea
\Omega \ket{D_{sgh}, \eta}^{(0)}_{\rm R} &=&
\delta({\gamma_0} -i\eta {\tilde \gamma}_0) 
\, \widetilde{\ket{-1/2}} \otimes  \ket{-1/2} \nonumber \\
&=&
-i \eta \, \delta({\tilde \gamma_0} +i\eta \gamma_0) 
\, \ket{-1/2} \otimes \widetilde{\ket{-1/2}} \nonumber \\
&=&  
-i \eta \, \ket{D_{sgh}, -\eta}^{(0)}_{\rm R} \, .
\eea
Furthermore, for the complete superghost state
\be
\ket{D_{sgh}, \eta}_{\rm R} =
{\rm exp}\left(i\eta \sum\limits_{n=1}^{\infty} 
(\gamma_{-n} {\tilde \beta}_{-n} - \beta_{-n} {\tilde \gamma}_{-n})
\right) \, \ket{D_{sgh}, \eta}_{\rm R}^{(0)} \, ,
\ee
using that $\Omega$ exchange left and right superghost oscillators,
we finally obtain:
\be
\Omega \ket{ D_{sgh}, \eta}_{\rm R} = -i \eta \ket{D_{sgh}, -\eta}_{\rm R}
\, .
\ee

Using the above result we can also derive
the $\O$-projection on the D-branes by using the
form of the R-R string state in the $(-1/2,-3/2)$ picture:
\bea
\ket{C^{(p+1)}} &=& {1 \over 2 (p+1)! \sqrt{2}} 
\, C^{(p+1)}_{\mu_1 \dots \mu_{p+1}}
\left( \left( {\cal C}_{(10)}\Gamma^{\mu_1 \dots \mu_{p+1}} \Pi_+ \right)_{AB}
\, {\rm cos} \gamma_0 \tilde{\beta}_0 \right. \\
&&\hspace{3cm}+ \,
\left. \left( {\cal C}_{(10)} \Gamma^{\mu_1 \dots \mu_{p+1}} \Pi_- \right)_{AB}
\, {\rm sin} \gamma_0 \tilde{\beta}_0 \right)
\, \ket{A}_{-1/2} \otimes \widetilde{\ket{B}}_{-3/2} \, .\nonumber
\eea
This analysis is very  useful for checking whether
there are spacetime-filling D-branes in the spectrum.
We have derived above that for the zero modes:
\be
\Omega \left( {\rm e}^{i\eta \gamma_0 {\tilde \beta}_0}
\ket{A}_{-1/2} \otimes \widetilde{\ket{B}}_{-3/2} \right)
= i\eta \, {\rm e}^{-i \eta \gamma_0 {\tilde \beta}_0} \,
\ket{B}_{-1/2} \otimes \widetilde{\ket{A}}_{-3/2} \, ,
\label{omega-groundstate-3}
\ee
which implies
\bea
\Omega \left( {\rm cos} \gamma_0 {\tilde \beta}_0 
\, \ket{A}_{-1/2} \otimes \widetilde{\ket{B}}_{-3/2} \right)
&=& {\rm sin} \gamma_0 {\tilde \beta}_0 
\, \ket{B}_{-1/2} \otimes \widetilde{\ket{A}}_{-3/2} \, ,\nonumber \\
\Omega \left( {\rm sin} \gamma_0 {\tilde \beta}_0 
\, \ket{A}_{-1/2} \otimes \widetilde{\ket{B}}_{-3/2} \right)
&=& {\rm cos} \gamma_0 {\tilde \beta}_0 
\, \ket{B}_{-1/2} \otimes \widetilde{\ket{A}}_{-3/2} \, .
\label{cos-sin-twist}
\eea
Finally, using this result and the properties of the 
$\Gamma$-matrices we find:
\be
\Omega \, \ket{C^{(p+1)}} = 
- (-1)^{{1 \over 2}p(p+1)} \, \ket{C^{(p+1)}} \, ,
\ee
hence it is invariant for $p=1,5,9$, as seen before.

Likewise, we can find  the action of $\Omega$ on the  quantum
state of the R-R twisted sector in the $(-1/2,-3/2)$ picture:
\bea
\ket{C^{(r+1)}}_{\rm T} &=& 
{1 \over \sqrt{2} (p+1)!} C^{(r+1)}_{\mu_1 \dots \mu_{r+1}} \, 
\left( \left( {\cal C}_{(6)}\gamma^{\mu_1 \dots \mu_{p+1}} \Pi_+ \right)_{ab}
\, {\rm cos} \gamma_0 \tilde{\beta}_0 
\right. \\
&&\hspace{3cm}+ \,
\left. \left( {\cal C}_{(6)}\gamma^{\mu_1 \dots \mu_{p+1}} \Pi_- \right)_{ab}
\, {\rm sin} \gamma_0 \tilde{\beta}_0 \right)
\, \ket{a}_{-1/2,{\rm T}} \otimes \widetilde{\ket{b}}_{-3/2,{\rm T}} \, .
\nonumber
\eea
Using (\ref{cos-sin-twist}) and the properties of the
6-dimensional $\gamma$-matrices we find:
\be
\O \, \ket{C^{(r+1)}}_{\rm T} =
(-1)^{{1 \over 2}r(r+1)} \, \ket{C^{(r+1)}}_{\rm T} \, ,
\ee
hence it is invariant for $r=-1,$ and $3$.

%%%%%%%%%%%%%%%%%%%%%%%%%%%%%%%%%%%%%%%%%%%%%%%%%%%%%%%%%%%%%%%%%%

  \section{Action of $\O$ on the Twisted NS-NS vacuum}
  \label{appendixA2}

In this Appendix we give an explicit analysis of the
action of $\Omega$ in the twisted NS-NS vacuum. This vacuum state consists
of the spin fields constructed out of the four fermionic zero modes in the
left and four from the right moving sector. The superghost vacuum is taken
to be in the $(-1,-1)$ picture. In each of the left and right sector, these
fermion zero modes satisfy an $SO(4)$ Clifford algebra. Thus we have
\be
\{\psi^i_0, \psi^j_0 \} = \delta^{ij} = \{\tilde {\psi}^i_0,
\tilde {\psi}^j_0 \} \, ,\qquad 
\{\psi^i_0, \tilde {\psi}^j_0 \} = 0.
\ee
where $i,j$ take values from 6,7,8 and 9. 
Our conventions for the $SO(4)$ gamma-matrices are as follows:
\be
\{\bar{\gamma}^i , \bar{\gamma}^j \} = 2 \delta^{ij} \, ,\qquad
{\cal C}_{(4)}^T = - {\cal C}_{(4)} \, ,\qquad 
[ {\cal C}_{(4)} , \bar{\gamma} ] = 0 \, ,\qquad 
(\bar{\gamma}^i)^T = - {\cal C}_{(4)}\bar{\gamma}^i 
{\cal C}_{(4)}^{-1} \, ,
\ee
with ${\bar \gamma} = - \bar{\gamma}^6 \bar{\gamma}^7 \bar{\gamma}^8 
\bar{\gamma}^9$.
Let $\alpha , \beta $ denote the spinor indices in four dimensions and let
$\ket{\alpha}_{\rm T}\otimes \widetilde{\ket{\beta}}_{\rm T}$ denote the
twisted spinor vacuum
constructed from the spin fields of the above fermionic zero modes. The
action of the fermionic oscillators and zero modes in this basis is
defined to be
\bea
\psi^i_n \ket{\alpha}_{\rm T}\otimes {\widetilde{\ket{\beta}}}_{\rm T} &=& 
\tilde{\psi}^i_n \ket{\alpha}_{\rm T}\otimes {\widetilde{\ket{\beta}}}_{\rm
T} = 0 \, ,\qquad
\forall n \geq 1 \, ,\nonumber\\
\psi^i_0 \ket{\alpha}_{\rm T}\otimes {\widetilde{\ket{\beta}}}_{\rm T} &=&
{1\over{\sqrt
2}} \left(\bar{\gamma}^i\right)^\alpha {}_\delta \,\left(\one \right)^\beta
{}_\rho \, \ket{\delta}_{\rm T}\otimes \widetilde{\ket{\rho}}_{\rm T} \,
,\nonumber\\
\tilde{\psi}^i_0 \ket{\alpha}_{\rm T}\otimes \widetilde{\ket{\beta}}_{\rm
T} &=& 
{1\over{\sqrt2}} \left(\bar{\gamma}\right)^\alpha {}_\delta
\,\left(\bar{\gamma}^i\right)^\beta{}_\rho \, \ket{\delta}_{\rm T}\otimes
\widetilde{\ket{\rho}}_{\rm T}\, .
\label{appendixA2-def1}
\eea
As in the earlier cases, the action of $\Omega$ is to interchange the left
and right moving sectors. Thus the defining relation for the action
of $\Omega$ can be obtained by demanding simply that, namely
\be
\Omega \left(\tilde{\psi}^i_0 \ket{\alpha}_{\rm T}\otimes 
\widetilde{\ket{\beta}}_{\rm T}\right) =  
{\psi}^i_0 
\widetilde{\ket{\alpha}}_{\rm T} 
\otimes 
{\ket{\beta}}_{\rm T} \, ,
\label{app-first-def}
\ee
or equivalently
\be
\Omega \left(\psi^i_0 \ket{\alpha}_{\rm T}\otimes 
\widetilde{\ket{\beta}}_{\rm T}\right) =  
\tilde{\psi}^i_0 
\widetilde{\ket{\alpha}}_{\rm T}
\otimes 
\ket{\beta}_{\rm T} \, ,
\label{app-second-def}
\ee
where
\be
\widetilde{\ket{\alpha}}_{\rm T}
\otimes 
\ket{\beta}_{\rm T} \, 
=
- \, \ket{\beta}_{\rm T} 
\otimes 
\widetilde{\ket{\alpha}}_{\rm T}
\ee 
These definitions guarantee that $\Omega^2 = 1$. Using the equations
in (\ref{appendixA2-def1}), the first equation (\ref{app-first-def}) gives:
\be
\left(\bar{\gamma} \right)^\alpha{}_\delta 
\,\left(\bar{\gamma}^i \right)^\beta{}_\rho 
\,\Omega \left(\ket{\delta}_{\rm T}\otimes 
\widetilde{\ket{\rho}}_{\rm T}\right) 
= - \,
\left(\bar{\gamma}^i \right)^\beta{}_\rho 
\, \left(\one \right)^\alpha {}_\delta
\, \ket{\rho}_{\rm T}\otimes
\widetilde{\ket{\delta}}_{\rm T}\, ,
\ee
which can be simplified to obtain
\be
\Omega \left(\ket{\alpha}_{\rm T}\otimes 
\widetilde{\ket{\beta}}_{\rm T}\right) =
- \, 
\left(\bar{\gamma}\right)^\alpha{}_\delta 
\, \ket {\beta}_{\rm T}
\otimes 
\widetilde{\ket{\delta}}_{\rm T} \, .
\ee
Similarly, the equation (\ref{app-second-def}) gives rise to
\be
\left(\bar{\gamma}^i\right)^\alpha{}_\delta 
\,\left(\one \right)^\beta{}_\rho 
\,\Omega \left(\ket{\delta}_{\rm T}\otimes 
\widetilde{\ket{\rho}}_{\rm T}\right) 
=
- \, 
 \left(\bar{\gamma}\right)^\beta{}_\rho 
\, \left(\bar{\gamma}^i\right)^\alpha {}_\delta
\, \ket{\rho}_{\rm T}\otimes
\widetilde{\ket{\delta}}_{\rm T}\, ,
\ee
which can be simplified to obtain
\be
\Omega \left(\ket{\alpha}_{\rm T}\otimes 
\widetilde{\ket{\beta}}_{\rm T}\right) =
- \,
\left(\bar{\gamma}\right)^\beta{}_\delta 
\, \ket {\delta}_{\rm T}
\otimes 
\widetilde{\ket{\alpha}}_{\rm T} \, .
\ee
If we denote the superghost vacuum in the $(-1, -1)$ picture as
$\ket{-1} \otimes \widetilde{\ket{-1}}$, with
\be
\Omega \left( \ket{-1} \otimes \widetilde{\ket{-1}} \right)
= \widetilde{\ket{-1}} \otimes {\ket{-1}} 
= - \ket{-1} \otimes \widetilde{\ket{-1}} \, ,
\ee
and noting that the full vacuum is given by
\be
\ket{\alpha}_{\rm -1,T} \otimes \widetilde{\ket{\beta}}_{\rm -1,T}
= \ket{\alpha}_{\rm T} \ket{-1} \otimes 
\widetilde{\ket{\beta}}_{\rm T} \widetilde{\ket{-1}} \, ,
\ee
we can write the action of $\Omega$ on the full vacuum 
of the twisted NS-NS sector as follows:
\be
\Omega \left( \ket{\alpha}_{\rm -1,T} \otimes \widetilde{\ket{\beta}}_{\rm
-1,T}\right)
= ({\bar{\gamma}})^{\alpha}{}_{\delta} \,
\ket{\beta}_{\rm -1,T} \otimes \widetilde{\ket{\delta}}_{\rm -1,T}
= ({\bar{\gamma}})^{\beta}{}_{\delta} \, 
{\ket{\delta}}_{\rm -1,T}  \otimes \widetilde{\ket{\alpha}}_{\rm -1,T}
\, ,
\ee
as previously announced in (\ref{(2.34)}).
Finally, note that this relation is equivalent to
\be
\ket{\alpha}_{\rm -1,T} \otimes \widetilde{\ket{\beta}}_{\rm -1,T}
= \left( \bar{\gamma} \right)^\alpha {}_\delta \,
\left( \bar{\gamma} \right)^\beta {}_\rho \,
\ket{\delta}_{\rm -1,T} \otimes \widetilde{\ket{\rho}}_{\rm -1,T} \, ,
\ee
which is consistent since both left and right 
spinors are of the same chirality.

  \section{Crosscaps in Type I on $T^4/\Z_2$}
  \label{appendixB}

The tadpole-cancelling D9 and D5 branes in type I on
$T^4/{\cal I}_4$ are bulk branes. 
In type I on a K3 orbifold we have a 9-crosscap and 5-crosscaps
which have boundary states in the untwisted sectors 
only:
\be
\ket{C9}=\ket{C9}_{\rm NS} + \ket{C9}_{\rm R} \, ,\qquad
\ket{C5}=\ket{C5}_{\rm NS} + \ket{C5}_{\rm R} \, .
\ee
As usual, each part is defined in terms of the spin structures:
\bea
\ket{C}_{\rm NS} &=& {1 \over 2} 
\left( \ket{C,+}_{\rm NS} - \ket{C,-}_{\rm NS} \right) \, ,\nonumber \\
\ket{C}_{\rm R} &=& {1 \over 2} 
\left( \ket{C,+}_{\rm R} + \ket{C,-}_{\rm R} \right) \, ,
\eea
where
\be
\ket{C,\eta}_{\rm NS/R} =
\ket{C_X} \ket{C_\psi, \eta}_{\rm NS/R} \ket{C_{gh}} 
\ket{C_{sgh}, \eta}_{\rm NS/R} \, .
\ee
This generic form is common to both crosscaps:
\bea
\ket{C_\psi, \eta}_{\rm NS} &=& \prod\limits_{r=1/2}^\infty
{\rm e}^{i \eta (-1)^r \psi_{-r} \cdot S \cdot \tilde\psi_{-r}} \,
\ket{0} \, ,\nonumber \\
\ket{C_{sgh},\eta}_{\rm NS} &=& \prod\limits_{r=1/2}^\infty
{\rm e}^{i \eta (-1)^r ( \gamma_{-r} \tilde\beta_{-r} -
\beta_{-r} \tilde\gamma_{-r} )} \, \ket{-1} \otimes \widetilde{\ket{-1}} \, ,
\nonumber \\
\ket{C_{gh}} &=& \prod\limits_{n=1}^\infty
{\rm e}^{ (-1)^n (c_{-n} \tilde{b}_{-n} - b_{-n} \tilde{c}_{-n})}
\left( {c_0 + \tilde{c}_0 \over 2} \right)\, 
\ket{1} \otimes \widetilde{\ket{1}}
\, ,\nonumber \\
\ket{C_\psi, \eta}_{\rm R} &=& \prod\limits_{m=1}^\infty
{\rm e}^{i \eta (-1)^m \psi_{-m} \cdot S \cdot \tilde\psi_{-m}} \,
\ket{C_\psi, \eta}^{(0)}_{\rm R} \, ,\nonumber \\
\ket{C_{sgh},\eta}_{\rm R} &=& \prod\limits_{m=1}^\infty
{\rm e}^{i \eta (-1)^m ( \gamma_{-m} \tilde\beta_{-m} -
\beta_{-m} \tilde\gamma_{-m} )} \, 
\ket{C_{sgh},\eta}^{(0)}_{\rm R}
\, ,
\eea
with $S_{\mu\nu} = \eta_{\mu\nu}$ for the 9-crosscap and
 $S_{\mu\nu} = (\eta_{\alpha\beta}, - \delta_{ij})$,
for the 5-crosscap, where
$\alpha,\beta= 0, \dots, 5$ are the directions on the fixed planes
and  $i,j=6,\dots,9$ are the directions along $T^4$.
The zero modes in the R-R sector are given by:
\bea
\ket{C_\psi, \eta}^{(0)}_{\rm R} &=& 
\left({\cal C}_{(10)} 
\Gamma^0 \cdots \Gamma^p {1 + i\eta \Gamma_{11} \over 1+i\eta}
\right)_{AB} \ket{A} \otimes \widetilde{\ket{B}} 
\, ,\qquad p=5,9 \, ,\nonumber \\
\ket{C_{sgh},\eta}^{(0)}_{\rm R} &=& {\rm e}^{i\eta \gamma_0 \tilde\beta_0}
\, \ket{-1/2} \otimes \widetilde{\ket{-3/2}} \, ,
\eea
The fact that the orbifold is compact is reflected in the
bosonic-oscillator part of the crosscaps states. 
The 9-crosscap contains windings along the $T^4$:
\be
\ket{C9_X} = {\cal N}_9 
\left( \sum\limits_{m \in \Z} {\rm e}^{i q_m {m R \over \alpha^\prime}}
\right)^4 \prod\limits_{n=1}^\infty 
{\rm e}^{- {1 \over n} (-1)^n \alpha_{-n} \cdot S \cdot \tilde\alpha_n}
\, \ket{0} \, ,
\ee
whereas the 5-crosscap contains momentum modes along the $T^4$ directions:
\be
\ket{C5_X} = {\cal N}_5 
\left( \sum\limits_{n \in \Z} {\rm e}^{i q_n {n \over R}}
\right)^4 \prod\limits_{n=1}^\infty 
{\rm e}^{- {1 \over n} (-1)^n \alpha_{-n} \cdot S \cdot \tilde\alpha_n}
\, \ket{0} \, .
\ee
For simplicity we have taken all four radii equal to each other.
The normalisation of these crosscaps
states is derived by comparing the open string 
M${\ddot {\rm o}}$bius amplitudes with a closed string calculation:
\bea
{\cal N}_9 &=& {1 \over \sqrt{2}}
\,  2^5 \, {T_9 \over 2} \left({2\pi R \over \Phi} \right)^2
\, , \nonumber \\
{\cal N}_5 &=& - \sqrt{2} \, 
\, {T_5 \over 2} \left({2\pi R \over \Phi} \right)^2
(2\pi R)^{-4} \, ,
%T_p &=& \sqrt{\pi} (2\pi \sqrt{\alpha^\prime})^{3-p} \, ,\qquad
%p=r+s \, ,\nonumber
\eea
where the relative sign between ${\cal N}_9$ and ${\cal N}_5$ stems
from the fact that $\Omega$ acts with a relative sign on the D9 and D5
branes \cite{Gimon} and $T_p$ is given in (\ref{tension-truncated}). 
%\bea
%N_p^\prime &=& m_p {T_p \over 2} \left({2\pi R \over \Phi} \right)^2
%(2\pi R)^{-4+s} \, , \nonumber \\
%{\cal N}_9 &=& n_9 {1 \over \sqrt{2}}
%\,  2^5 \, {T_9 \over 2} \left({2\pi R \over \Phi} \right)^2
%\, , \nonumber \\
%{\cal N}_5 &=& n_5 {1 \over \sqrt{2}} \, 
%2^5 \,  {T_5 \over 2} \left({2\pi R \over \Phi} \right)^2
%(2\pi R)^{-4} \, , \nonumber \\
%T_p &=& \sqrt{\pi} (2\pi \sqrt{\alpha^\prime})^{3-p} \, ,\qquad
%p=r+s \, .\nonumber
%\eea

\end{document}